\documentclass[12pt,singlespacing]{elsarticle}
\usepackage{graphicx}
\usepackage{amsmath}
\usepackage[table]{xcolor}
\usepackage{multirow}
\usepackage{amssymb} 
\usepackage{setspace}
\usepackage{times}
\usepackage{soul}
\usepackage{fullpage}
\usepackage{datetime}
\usepackage{caption}
\usepackage{subcaption}
\usepackage{placeins}
\usepackage{bm}
\usepackage{bm,upgreek}
\usepackage{amsbsy}
\usepackage{algorithm}
\usepackage[normalem]{ulem}
\usepackage{natbib}
\usepackage{comment}
\usepackage{ulem}

\makeatletter
\def\ps@pprintTitle{
 \let\@oddhead\@empty
 \let\@evenhead\@empty
 \def\@oddfoot{}
 \let\@evenfoot\@oddfoot}
\makeatother

\newcommand{\pderiv}[2]{\dfrac{\partial #1}{\partial #2}}

\newtheorem{remark}{Remark}
\newcommand{\eref}[1]{(\ref{#1})}

\newcommand{\jump}[1]{\ensuremath{[\![#1]\!]} }

\begin{document}

\begin{frontmatter}
\title{A stabilized finite element method for delamination analysis of composites using cohesive elements} 

\author[vu]{Gourab Ghosh}
\author[vu]{Ravindra Duddu\corref{cor1}}
\ead{ravindra.duddu@vanderbilt.edu}
\author[ll]{Chandrasekhar Annavarapu}

\cortext[cor1]{Corresponding author}
\address[vu]{Department of Civil and Environmental Engineering, Vanderbilt University, Nashville, Tennessee.}
\address[ll]{Department of Civil Engineering, Indian Institute of Technology Madras, Chennai, India.
 \vspace{-5mm}}

\begin{abstract}
We demonstrate the ability of a stabilized finite element method, inspired by the weighted Nitsche approach, to alleviate spurious traction oscillations at interlaminar interfaces in multi-ply multi-directional composite laminates. In contrast with the standard (penalty-like) method, the stabilized method allows the use of arbitrarily large values of cohesive stiffness and obviates the need for engineering approaches to estimate minimum cohesive stiffness necessary for accurate delamination analysis. This is achieved by defining a weighted interface traction in the stabilized method, which allows a gradual transition from penalty-like method for soft elastic contact to Nitsche-like method for rigid contact. We conducted several simulation studies involving constant strain patch tests and benchmark delamination tests under mode-I, mode-II and mixed-mode loadings. Our results show clear evidence of traction oscillations with the standard method with structured and perturbed finite element meshes, and that the stabilized method alleviates these oscillations, thus illustrating its robustness.
\end{abstract}


\end{frontmatter}

\section{Introduction}\label{Introduction}

In laminated fiber-reinforced composites, delamination is one of the most dominant failure mechanisms, which involves progressive damage accumulation and fracture along interlaminar interfaces \citep{mi1998progressive}. Delamination under static and cyclic-fatigue loading has been widely studied in the literature over the past 40 years \citep{pascoe2013methods,banks201450th}, because it causes localized damage that is hard to detect and may lead to sudden structural collapse. The cohesive zone modeling approach has been extensively used to analyze and predict mixed-mode delamination propagation, despite its drawbacks and limitations. A particular drawback of the standard finite element implementation of cohesive zone models (CZMs) is its occasional numerical instability, which causes spurious traction oscillations at delamination/crack interface \citep{schellekens1993non}. Simple engineering solutions \citep{Turon2007} may mitigate numerical issues with the standard finite element method (FEM) on a case-by-case basis, but they are not robust and introduce parametric uncertainty  \citep{jimenez2016parametric}. We recently illustrated that a Nistche-based stabilized FEM is robust and accurate for enforcing stiff cohesive laws and simulating fracture propagation in isotropic, homogeneous elastic materials \citep{ghosh2019stabilized}. The purpose of this article is to investigate the numerical stability and accuracy of standard FEM and weighted Nitsche-based stabilized FEM for delamination analysis of composites using cohesive elements, especially at anisotropic and dissimilar interlaminar interfaces.

The cohesive zone modeling approach, which is based on continuum damage mechanics \citep{alfano2001finite}, has been widely used to simulate mixed-mode delamination of composites \citep{camanho2003numerical,jiang2007concise,turon2010accurate}, including the growth of multiple delamination cracks extending into various ply interfaces \citep{may2010combined,higuchi2017numerical}. Despite its success, the standard FEM implementation of CZMs comes with certain outstanding challenges, including mesh dependence, parametric uncertainty, computational efficiency, and numerical instability. Mesh dependence of the predicted crack path or directional mesh bias can be an issue with CZMs, as cracks can only propagate along finite element edges. To allow the propagation of arbitrary cohesive cracks and/or the inclusion of intra-element cohesive interfaces, various approaches based on the partition of unity concept (including G/XFEM) or virtual/phantom nodes were proposed \citep{wells2001new,moes2002extended,remmers2003cohesive,mergheim2005finite,Song2006Phantom,maiti2009generalized,zhang2019igfem,zhi2019geometrically}. The performance of the standard FEM to simulate cohesive cracks can be poor with distorted or low quality meshes, so mesh free methods were proposed to alleviate such difficulties \citep{rabczuk2007meshfree}. Recently, phase-field damage models have also been proposed to simulate delamination of orthotropic laminates \citep{dhas2018phase,denli2020phase}. 

Parametric uncertainty or sensitivity of CZMs to the choice of model parameters can affect the accuracy of delamination analysis. Depending on the choice of cohesive stiffness, cohesive/interface strength, and fracture toughness parameters, the accuracy of load and interface traction prediction can vary significantly \citep{jimenez2016parametric,Lu2019cohesive}. Out of these three parameters, only fracture toughness can be determined or well-constrained from experiments for a given composite material. Notably, in the case of mixed-mode delamination at dissimilar interfaces in bi-materials, fracture toughness is also dependent on the mode mixity (loading) \citep{freed2008new}. In contrast, the cohesive strength is difficult to determine from experiments, and is usually assumed based on mesh size considerations \citep{alfano2001finite,Turon2007,harper2008cohesive}, as a compromise between computational efficiency and numerical accuracy. The cohesive stiffness is generally regarded as a numerical penalty parameter in intrinsic CZMs that assume an initially elastic response, and in extrinsic CZMs that assume initially rigid response and elastic unloading/reloading response. The FEM implementation of extrinsic CZMs requires advanced algorithms \citep{mota2008fracture,espinha2013scalable,settgast2017fully}, which increases computational complexity; whereas, intrinsic CZMs are relatively straightforward to implement within a legacy/commercial finite element code, but the choice of cohesive stiffness can affect numerical stability and/or convergence. It is noteworthy that extrinsic CZMs under fatigue and compressive loading scenarios also suffer from numerical instabilities observed in intrinsic CZMs with large values of cohesive stiffness.

In quasi-static fracture/delamination analysis, the standard FEM implementations using intrinsic CZMs can exhibit spurious traction oscillations along the cohesive interface, especially near crack tips, if a large initial cohesive stiffness is specified \citep{schellekens1993numerical,elices2002cohesive,de2003numerical,simone2004partition,yang2005cohesive,Turon2007,thai2017numerical}. Even with potential-based intrinsic CZMs, it is important to control the elastic behavior through initial slope indicators to avoid instability  \citep{park2009unified}. Past studies indicate that using the Gaussian full integration scheme \citep{schellekens1993non,schellekens1993numerical,mi1998progressive,de2003numerical} can cause spurious oscillations in the traction profile along cohesive interfaces. Although the Newton-Cotes integration scheme was suggested as an alternative, some studies reported that it can result in an odd wrinkling mode in thin laminates undergoing delamination process \citep{davila2008effective,davila2009procedure,yang2010improved}. Improved cohesive stress integration schemes for continuum CZMs \citep{do2013improved} were developed to further tackle the issues of stability and robustness. Alternatively, discrete CZMs were developed  \citep{cui1993combined,xie2006discrete,liu2012discrete,wang2015progressive} that use spring-like elements to connect the finite element nodes at delamination interfaces, instead of element edges, to alleviate numerical instability and/or convergence issues. However, in discrete CZMs, when using non-uniform meshes, or modeling interfacial kinks, relating force-displacement relations for the spring like elements with interface traction is not straightforward. The issue of numerical instability in the standard FEM implementation of intrinsic continuum CZMs, when using large cohesive stiffness and full/reduced Gauss integration schemes, arises due to ill-conditioning of discrete systems, as typical with penalty-like formulations \citep{svenning2016weak}. Although Lagrange-multiplier-based mixed or two-field formulations for intrinsic/extrinsic CZMs  \citep{lorentz2008mixed,cazes2013two} can overcome numerical instability, they can be computationally expensive or cumbersome owing to the difficulty in determining a stable Lagrange multiplier space.

To broadly address the numerical instability issues with penalty-like formulations and standard FEM for interface/contact problems, discontinuous Galerkin (dG) methods or Nitsche-based methods were proposed in the last two decades \citep{hansbo2005nitsche,nguyen2014discontinuous}. Several novel dG approaches were developed for fracture problems, including dG interface \citep{truster2013discontinuous,Versino2015,Aduloju2019}, space-time dG approaches \citep{Abedi2018,abedi2019computational}, and hybrid dG-CZM approaches \citep{mergheim2004hybrid,seagraves2015large,nguyen2016modelling,bayat2019}, where the dG method is generally used before fracture initiation and CZMs are used for simulating fracture propagation. A limitation, however, for the wider use of dG approaches for fracture is the inherent complexity associated with their implementation, especially in legacy/commercial finite element codes/software. Conversely, Nitsche methods were advocated by \citep{hansbo2004finite,sanders2009methods} for modeling strong/weak discontinuities and elastic interface problems. Nitsche methods overcome the issue of numerical instability affecting penalty-like formulations, by adding consistency terms \citep{liu2008contact,wriggers2008formulation}. Nitsche's method has been extended for modeling frictional-sliding on
embedded interfaces \citep{Annavarapu2014part2,annavarapu2014nitsche} and small-sliding contact on frictional surfaces, including stick?slip behavior \citep{annavarapu2015weighted}. Inspired by the work of \citep{Juntunen2009}, we recently extended the Nitsche's method to cohesive fracture problems, and developed a stabilized FEM that alleviates traction oscillations with stiff, anisotropic cohesive laws \citep{ghosh2019stabilized}.

In this paper, we illustrate the ability of the stabilized FEM of Ghosh \textit{et al.} \citep{ghosh2019stabilized} in alleviating traction oscillations at interlaminar interfaces in multi-directional orthotropic composite laminates under different loading conditions. A specific aim is to illustrate its robustness for composite delamination analysis, with regard to the choice of the cohesive stiffness and the structure of the finite element mesh (e.g. uniform structured versus perturbed or semi-structured meshes), which has not been addressed before. The rest of this paper is organized as follows: in Section~\ref{sec:MoForm}, we briefly describe the governing equations of the cohesive fracture problem and the weak forms corresponding to the standard and stabilized methods. In Section~\ref{sec:NumImp}, we discuss the salient aspects of the numerical implementation and the selected model/material parameters. Specifically, we focus on the implementation via user element subroutines in commercial finite element software ABAQUS \citep{ABAQUSManual}, so that it can be utilized by the broader composite modeling community. In Section~\ref{sec:NumEx}, we present several benchmark numerical examples to compare the standard and stabilized methods, with a particular emphasis on the accuracy of the interface traction field and load--displacement curves. Finally, in Section~\ref{sec:Conc}, we conclude with a summary and closing remarks.

\section{Governing equations and weak formulations} \label{sec:MoForm}

In this section, we briefly review the Nitsche-inspired stabilized finite element method originally proposed in \cite{ghosh2019stabilized} for enforcing stiff cohesive laws. We will begin with a description of the strong form of the governing equations followed by the anisotropic bilinear cohesive law for mixed-mode loading. Subsequently, we will discuss the weak form for the standard and stabilized methods, and the choice of stabilization parameters and weights. 

\subsection{Strong form of the delamination/debonding problem} \label{Governing Equations}
\begin{figure}[ht]
\centering
\includegraphics[width=0.35\textwidth]{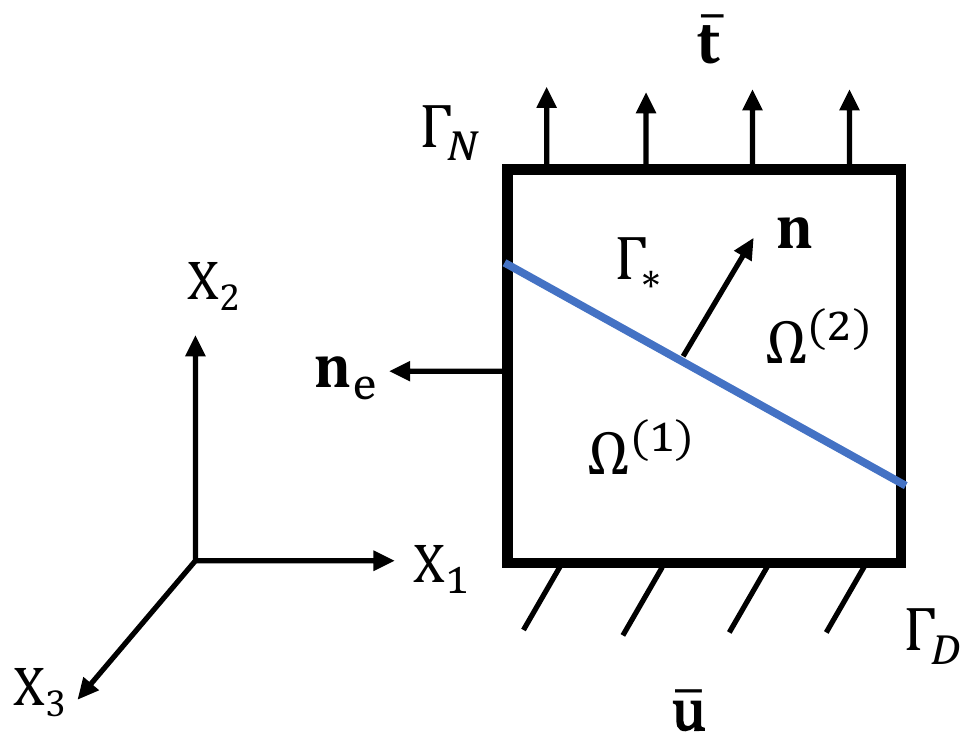}
\caption{A schematic of the undeformed domain for the quasi-static delamination/debonding problem. We choose X$_1$ and X$_2$ as the in-plane coordinates for the laminate material, and X$_3$ is the out-of-plane coordinate.}
\label{fig:domain}
\end{figure}

We define an initial domain $\Omega$ $\subset$ $\mathbb{R}^{2}$, which is partitioned into two non-overlapping bulk domains $\Omega^{(1)}$ and $\Omega^{(2)}$ separated by a pre-defined internal cohesive interface $\Gamma_{*}$, such that $\Omega$ = $\Omega^{(1)}$ $\cup$ $\Omega^{(2)}$ (see Fig. \ref{fig:domain}). Throughout this paper, we use the notation that numbers within parentheses in the superscript identify the domain partitions. Dirichlet and Neumann boundary conditions are enforced on two disjointed parts of the domain boundary $\Gamma\equiv\partial\Omega$ in such a way so that $\partial\Omega=\Gamma_{D}\cup\Gamma_{N}$ with $\Gamma_{D}\cap \Gamma_{N}=\emptyset$. The outward unit normal to the boundary $\partial\Omega$ is denoted by $\mathbf{n}_{\mathrm{e}}$ and unit normal vector associated with the interface boundary $\Gamma_{*}$ is denoted by $\mathbf{n}$ and points from $\Omega^{(2)}$ to $\Omega^{(1)}$ (thus $\mathbf{n}=-\mathbf{n}^{(1)}=\mathbf{n}^{(2)}$). 

For delamination and debonding at sharp interfaces, we assume the bulk domains consist of a homogeneous anisotropic linearly elastic material, thus intralaminar damage is neglected and only interlaminar damage is considered. Assuming small displacements, the Cauchy stress tensor can be defined in $\Omega^{(1)}$ and $\Omega^{(2)}$ as
\begin{equation}
\boldsymbol{\sigma}^{(m)}=\mathbf{D}^{(m)}:\boldsymbol{\epsilon}^{(m)}, ~~~ m=\{1,2\},
\label{cauchystressexpression}
\end{equation}
where $\mathbf{D}$ denotes the fourth-order anisotropic elasticity tensor and the small strain tensor $\boldsymbol{\epsilon}
= \frac{1}{2}(\boldsymbol\nabla\mathbf{u}+(\boldsymbol\nabla\mathbf{u})^{T})$ is defined by the symmetric part of the displacement gradient tensor. The governing elasto-static equilibrium equations in the absence of body forces are given by:
\begin{eqnarray}
\boldsymbol{\nabla }\cdot\boldsymbol{\sigma}^{(m)} &=& \mathbf{0} ~~\text{in}~~ \Omega^{(m)}, ~~~ m=\{1,2\}, \label{eq:cauchycohesive} \\
\mathbf{u} &=& \bar{\mathbf{u}} ~~\text{on}~~ \Gamma_{D}, \label{eq:Dirichlet} \\
\boldsymbol{\sigma}\cdot \mathbf{n}_{\text{e}}&=& \bar{\mathbf{t}} ~~\text{on}~~ \Gamma_{N},  \label{eq:Neumann} \\
\mathbf{t}_{\text{c}} &=& - \boldsymbol{\sigma}^{(1)}\cdot\mathbf{n}^{(1)} = \boldsymbol{\sigma}^{(2)} \cdot\mathbf{n}^{(2)}
~~\text{on}~~ \Gamma_{*},  
\label{eq:tractioncont}
\end{eqnarray}
where $\bar{\mathbf{u}}$ is the prescribed displacement vector on the Dirichlet boundary $\Gamma_{D}$ and $\bar{\mathbf{t}}$ is the prescribed traction on the Neumann boundary $\Gamma_{N}$. The interface traction $\mathbf{t}_{\text{c}}$ is related to the Cauchy stress tensor evaluated in the sub-domains $\Omega^{(1)}$ and $\Omega^{(2)}$ as given by \eref{eq:tractioncont}, and is continuous across the cohesive interface $\Gamma_*$ to satisfy the force equilibrium. 
The cohesive traction $\hat{\mathbf{t}}_{\text{c}}$ is the Newton's third law pair to the interface traction $\mathbf{t}_{\text{c}}$ on a given delamination/crack surface and can be defined as a function of the interface separation or displacement jump as
\begin{equation}
\hat{\mathbf{t}}_{\text{c}} = - \mathbf{t}_{\text{c}} = \boldsymbol{\alpha}(\boldsymbol{\delta})~\boldsymbol{\delta}, 
\label{eq:TSRelation}
\end{equation}
\begin{equation}
\boldsymbol{\delta}=\jump{\mathbf{u}}=\mathbf{u}^{(2)}-\mathbf{u}^{(1)},
\label{eq:dispjump}
\end{equation} 
where the cohesive stiffness matrix $\boldsymbol{\alpha}$ is usually a nonlinear function of the interface separation. 

\begin{remark}
 In the case where the interlaminar interface delineates two dissimilar materials (i.e., $\mathbf{D}^{(1)} \neq \mathbf{D}^{(2)}$), the Cauchy stress tensor evaluated at interface from either side can be different (i.e., $\boldsymbol{\sigma}^{(1)} \neq \boldsymbol{\sigma}^{(2)}$), but the traction field must be continuous across the interface.
\end{remark}

\subsection{Anisotropic bilinear cohesive law} \label{IntrinsicCohesiveLaw}

\begin{figure}[ht]
 \centering
 \begin{subfigure}[b]{0.49\linewidth}
   \centering\includegraphics[width=\textwidth]{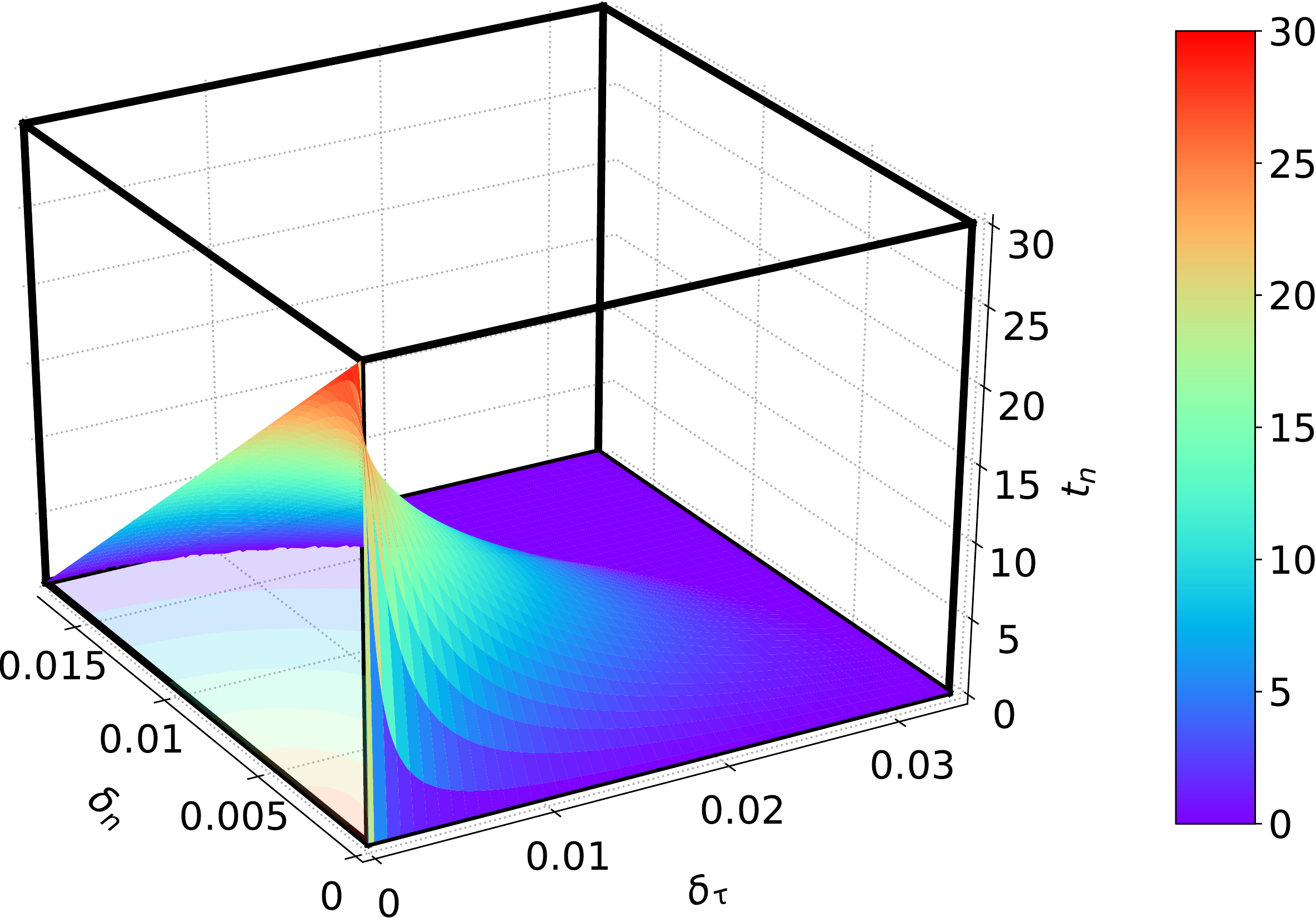}
   \caption{}
       \label{fig:normaltraction}
 \end{subfigure}
 ~
 \begin{subfigure}[b]{0.49\linewidth}
   \centering\includegraphics[width=\textwidth]{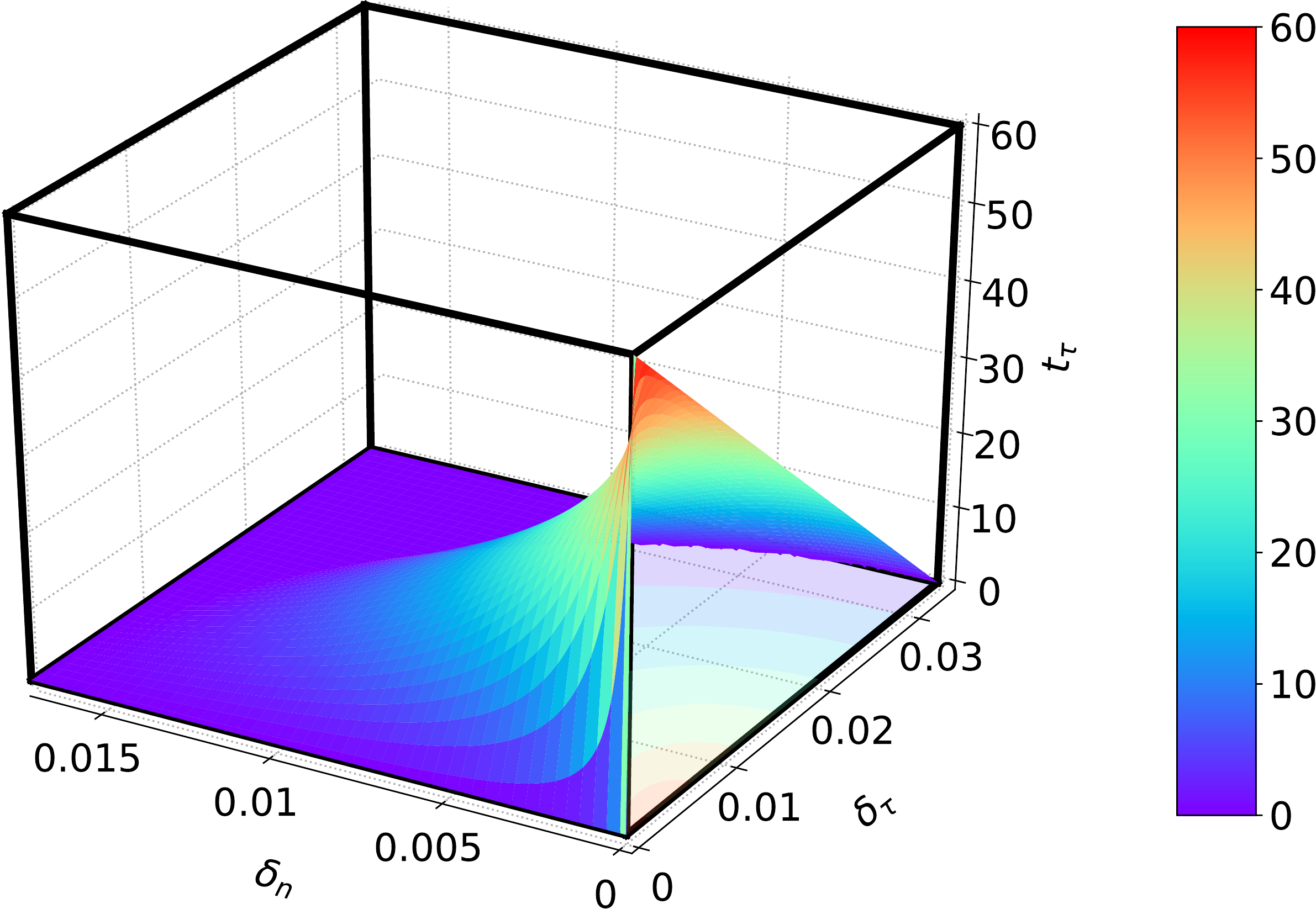}
   \caption{}
       \label{fig:tangentialtraction}
 \end{subfigure}
 
  \caption{Traction-separation relations defined by the anisotropic, intrinsic bilinear cohesive law of \citep{jiang2007concise}: (a) normal traction field and (b) tangential traction field.} \label{fig:cohesivelaw}
\end{figure}

We consider an intrinsic bilinear traction--separation or cohesive law that consists of an initial elastic region followed by a softening region. For mixed-mode delamination under quasi-static loading in two-dimensions, we cast the bilinear cohesive law in the damage mechanics framework as detailed in \cite{jimenez2016parametric}. The tangential $t_{\tau}$ and normal $t_{n}$ components of the interface traction vector $\mathbf{t}_{\text{c}}$ are related to the tangential $\delta_{\tau}$ and normal $\delta_{n}$ components of the interface separation  $\boldsymbol{\delta}$ as
\begin{equation}
\mathbf{t}_{\text{c}} = \begin{Bmatrix}
t_{\tau}\\
t_{n} 
\end{Bmatrix} = -\begin{bmatrix}
(1-d_{s}) \alpha_{\tau}^0 & 0\\ 
 0& \left(1-d_{s} \dfrac{\langle \delta_{n} \rangle}{\delta_{n}}\right) \alpha_n^0
\end{bmatrix}\begin{Bmatrix}
\delta_{\tau}\\ 
\delta_{n}
\end{Bmatrix}, \label{eq:BilinearTSLaw}
\end{equation}
\noindent where $\alpha_{n}^{0}$ and $\alpha_{\tau}^{0}$ represent the initial cohesive stiffness in the normal and the tangential directions, respectively, and the scalar damage variable $d_{s}$ is given by 
\begin{eqnarray} 
d_{s}=\left\{\begin{matrix}
&0 &\text{if} &\delta_{e}<\delta^{c}_{e},\\ 
&\dfrac{\delta^{u}_{e}(\delta_{e}-\delta^{c}_{e})}{\delta_{e}(\delta^{u}_{e}-\delta^{c}_{e})} & \text{if} & \delta^{c}_{e} \leq \delta_{e}<\delta^{u}_{e}, \\ 
&1 &\text{if} &\delta^{u}_{e} \leq \delta_{e},
\end{matrix}\right.
\label{eq:damagevariable}
\end{eqnarray}
and the equivalent separation $\delta_{e}=\sqrt{\langle\delta_{n}\rangle^{2}+\delta^{2}_{\tau}}$. In the above equations, $\langle \cdot \rangle$ denotes Macaulay brackets, so that $\langle \delta_{n} \rangle = \text{max}(0,\delta_{n})$, which ensures that there is no damage growth or damage effect on the normal cohesive stiffness response under compression or contact. The critical and ultimate interface separation parameters $\delta^{c}_{e}$ and $\delta^{u}_{e}$, respectively, defined as \citep{jiang2007concise}:
\begin{eqnarray}
&\dfrac{1}{\delta^{c}_{e}}=\sqrt{\left(\dfrac{\alpha^{0}_{n} \cos I}{\sigma_{\textrm{max}}}\right)^{2}+ \left(\dfrac{\alpha^{0}_{\tau} \cos II}{\tau_{\textrm{max}}}\right)^{2}},\\
&\dfrac{1}{\delta^{u}_{e}}=\left(\dfrac{\alpha^{0}_{n}\delta^{c}_{e}(\cos I)^{2}}{2~G_{IC}}\right)+\left(\dfrac{\alpha^{0}_{\tau}\delta^{c}_{e}(\cos II)^{2}}{2~G_{IIC}}\right),
 \label{eq:separationexpressions}
\end{eqnarray}
where the direction cosines $\cos I = \delta_{n}/\delta_{e}$ and $\cos II= \delta_{\tau}/ \delta_{e}$, $\sigma_{\textrm{max}}$ and $\tau_{\textrm{max}}$ are the pure mode I and mode II cohesive strengths, and $G_{IC}$ and $G_{IIC}$ are the pure mode I and mode II critical fracture energies. For illustration, the normal and tangential traction profiles as a function of the normal and tangential interface separations are shown in Fig. \ref{fig:cohesivelaw}.

\begin{remark}
Although the anisotropic bilinear cohesive law is phenomenological, it can be related to a potential function defined as
\begin{equation}
\Psi = - \dfrac{1}{2} \left( (1-d_{s}) \alpha_{\tau}^0 \delta_{\tau}^2 + \left(1-d_{s} \dfrac{\langle \delta_{n} \rangle}{\delta_{n}}\right)  \alpha_{n}^0 \delta_{n}^2 \right).
\end{equation}
The crack surface traction components can be defined based on the above potential function as $t_{\tau} = \dfrac{\partial \Psi}{\partial \delta_{\tau}}$ and $ t_{n} = \dfrac{\partial \Psi}{\partial \delta_{n}}$. The expressions for \eref{eq:BilinearTSLaw} are approximations obtained by neglecting the nonlinearity due to interface damage $d_s$, which is a function of the interface separation.  
\end{remark}

\begin{remark}
The anisotropic bilinear cohesive law of Jiang et al. \citep{jiang2007concise} has six independent parameters, namely initial cohesive stiffness, maximum cohesive strength and critical fracture energy of pure mode I and II loadings, to describe the traction-separation relationship. If these parameter values are chosen to be the same for both normal and shear modes, then we get an isotropic bilinear cohesive law with only three independent parameters. 
\end{remark}

\subsection{Standard weak form} \label{TradWeakForm}
\noindent We apply the Galerkin procedure of weighted residuals to derive the standard weak form, which is detailed in \cite{ghosh2019stabilized}. By weighting the equilibrium equation in (\ref{eq:cauchycohesive}) by a test function $\mathbf{w}$, integrating by parts, applying the divergence theorem, using the traction continuity condition at the interface in (\ref{eq:tractioncont}), and the constitutive relation in (\ref{cauchystressexpression}), we obtain the weak form as:
\begin{equation}
\sum_{m=1}^{2} \int_{\Omega^{(m)}}\nabla^{\text{s}}\mathbf{w}^{(m)}:\mathbf{D}^{(m)}:\nabla^{\text{s}}\mathbf{u}^{(m)} ~\text{d}\Omega - \int_{\Gamma_{*}}\jump{\mathbf{w}}\cdot\mathbf{t}_{\text{c}}~\text{d}\Gamma = \int_{\Gamma_{N}}\mathbf{w}\cdot\bar{\mathbf{t}}~\text{d}\Gamma, \label{eq:weakform(a)}
\end{equation}
where the jump in the test function is defined as $\jump{\mathbf{w}}=\mathbf{w}^{(2)}-\mathbf{w}^{(1)}$.
Substituting the traction-separation relation in \eqref{eq:TSRelation} into the weak form in \eqref{eq:weakform(a)} we obtain the standard weak form as

\begin{align}
&\sum_{m=1}^{2} \int_{\Omega^{(m)}}\nabla^{\text{s}}\mathbf{w}^{(m)}:\mathbf{D}^{(m)}:\nabla^{\text{s}}\mathbf{u}^{(m)} ~\text{d}\Omega + \int_{\Gamma_{*}} \jump{\mathbf{w}} \cdot ~ \boldsymbol{\alpha}(\boldsymbol{\delta})~\boldsymbol{\delta} ~\text{d}\Gamma =
\int_{\Gamma_{N}}\mathbf{w}\cdot \bar{\mathbf{t}}~\text{d}\Gamma.
\label{eq:weakform(b)}
\end{align}
Because the cohesive traction and separation components are defined in the normal and tangential directions, the standard weak form is implemented as, 

\begin{align}
&\sum_{m=1}^{2} \int_{\Omega^{(m)}}\nabla^{\text{s}}\mathbf{w}^{(m)}:\mathbf{D}^{(m)}:\nabla^{\text{s}}\mathbf{u}^{(m)} ~\text{d}\Omega + \int_{\Gamma_{*}} (1-d_{s}) \left(\jump{w_n} \alpha^{0}_{n} \delta_n + \jump{w_{\tau}} \alpha^{0}_{\tau} \delta_{\tau} \right) ~\text{d}\Gamma =
\int_{\Gamma_{N}}\mathbf{w}\cdot \bar{\mathbf{t}}~\text{d}\Gamma. \label{eq:weakform(c)}
\end{align} 

Thus, in the standard weak formulation, the cohesive traction is simply enforced as a mixed boundary condition on the interface. 

\begin{remark}
If the initial cohesive stiffness parameters $\alpha^{0}_{n}$ and $\alpha^{0}_{\tau}$ are taken to be large enough, the standard weak form resembles the penalty method for enforcing displacement continuity across the interface. However, for stiff cohesive laws, where cohesive stiffness is several orders of magnitude greater than the elastic modulus, the standard weak form becomes ill-conditioned leading to numerical instability and/or convergence issues. In the limiting case of a non-interpenetration (contact) constraint or an extrinsic cohesive law, where $\alpha^{0}_{n}\rightarrow \infty$ and/or $\alpha^{0}_{\tau}\rightarrow \infty$, the standard weak form is not well defined.
\end{remark}

\subsection{Stabilized weak form} \label{StabWeakForm}

The stabilized FEM developed in \cite{ghosh2019stabilized} extended Nitsche's method  \cite{Juntunen2009,annavarapu2012stable} to cohesive fracture problems, which we review here for the sake of clarity. The key idea is to evaluate the interface traction in terms of a weighted average stress in the bulk material across the interface and the traction in the cohesive interface as
\begin{equation}
  \mathbf{t}_{\text{c}} = (\mathbf{I}-\mathbf{S}) \left<\boldsymbol{\sigma}\right>_{\gamma} \cdot \mathbf{n} - \mathbf{S} \boldsymbol{\alpha}~\boldsymbol{\delta}, \label{eq:TracDef}
\end{equation}
where $\mathbf{I}$ is the second-order identity matrix, $\mathbf{S}$ is the stabilization matrix defined as, 
\begin{equation}
\mathbf{S}=\begin{bmatrix}
\dfrac{\beta_{\tau}}{\alpha^{0}_{\tau}(1-d_{s})+\beta_{\tau}} & 0\\
0 & \dfrac{\beta_{n}}{\alpha^{0}_{n}(1-d_{s})+\beta_{n}}
\end{bmatrix},
\label{eq:stabilization matrix}
\end{equation}
$\beta_{\tau}, \beta_{n}$ are the stabilization parameters, and $\left<\boldsymbol{\sigma}\right>_{\gamma}$ is the weighted average of the stress tensors on both sides of the interface defined as
\begin{equation}
\left<\boldsymbol{\sigma}\right>_{\gamma} = 
(\gamma^{(1)}\boldsymbol{\sigma}^{(1)}+\gamma^{(2)}\boldsymbol{\sigma}^{(2)}) 
~~ \forall ~~ \gamma^{(1)}+\gamma^{(2)}=1, ~~ \gamma^{(1)}>0, ~~ \gamma^{(2)}>0.
 \label{eq:weightsinNitsche}
 \end{equation}
Substituting \eref{eq:TracDef} into the weak form in \eref{eq:weakform(a)} we obtained the stabilized weak form as 

\begin{align}
\sum_{m=1}^{2} \int_{\Omega^{(m)}}\nabla^{\text{s}}\mathbf{w}^{(m)}:\mathbf{D}^{(m)}:\nabla^{\text{s}}\mathbf{u}^{(m)}  ~\text{d}\Omega &- \int_{\Gamma_{*}} \jump{\mathbf{w}} \cdot (\mathbf{I}-\mathbf{S})\left<\boldsymbol{\sigma}\right>_{\gamma} \cdot \mathbf{n}~\text{d}\Gamma \nonumber \\
&+ \int_{\Gamma_{*}} \jump{\mathbf{w}}\cdot\mathbf{S} \boldsymbol{\alpha}(\delta)~\boldsymbol{\delta} ~\text{d}\Gamma = \int_{\Gamma_{N}} \mathbf{w}\cdot \bar{\mathbf{t}}~\text{d}\Gamma.
 \label{eq:stabilizedweakform}
 \end{align}
In the above equation, the second and third terms on the left hand side ensure consistency and stability, respectively. If the stabilization matrix is taken as the identity matrix, then the stabilized weak form in \eref{eq:stabilizedweakform} becomes identical to the standard weak form in \eref{eq:weakform(b)}. 

\begin{remark}
The stabilized method presented here is unsymmetric and resembles the incomplete interior penalty method \citep{arnold2002unified,liu2009three}. It can be proved that the displacement solution $\mathbf{u}$ of the strong form equations \eref{eq:cauchycohesive} -- \eref{eq:tractioncont} is satisfied by the solution to the weak form equation \eref{eq:stabilizedweakform}, which establishes consistency for any value of cohesive stiffness; the mathematical procedure for proving this is similar to that described in \citep[][Lemma 2.1]{Juntunen2009}.
\end{remark}

The stabilized weak form with the interface traction and separation components expressed in the normal and tangential coordinates is given by 

\begin{multline}
\sum_{m=1}^{2} \int_{\Omega^{(m)}}\nabla^{\text{s}}\mathbf{w}^{(m)}:\mathbf{D}^{(m)}:\nabla^{\text{s}}\mathbf{u}^{(m)} ~\text{d}\Omega~ - \int_{\Gamma_{*}} \jump{\mathbf{w}} \cdot (\mathbf{I}-\mathbf{S})\left<\boldsymbol{\sigma}\right>_{\gamma} \cdot \mathbf{n}~\text{d}\Gamma~ \\
+ \int_{\Gamma_{*}} \left(\jump{w_n} \dfrac{(1-d_{s}) \alpha^{0}_{n}\beta_{n}}{\alpha^{0}_{n}(1-d_{s})+\beta_{n}} \delta_n + \jump{w_{\tau}} \dfrac{(1-d_{s}) \alpha^{0}_{\tau}\beta_{\tau}}{\alpha^{0}_{\tau}(1-d_{s})+\beta_{\tau}} \delta_{\tau} \right) ~\text{d}\Gamma =
\int_{\Gamma_{N}}\mathbf{w}\cdot \bar{\mathbf{t}}~\text{d}\Gamma. \label{eq:weakform(d)}
\end{multline}

\begin{remark}
As $(1-d_{s})\alpha^{0}_{n}, (1-d_{s})\alpha^{0}_{\tau} \rightarrow \infty$, the stabilized weak form in \eref{eq:weakform(d)} resembles the Nitsche stabilized finite element method for frictional contact presented in \citep{annavarapu2014nitsche}. Thus, the stabilized weak form remains well-defined in the limiting case of a non-interpenetration (contact) constraint or an extrinsic cohesive law, unlike the standard weak form in \eref{eq:weakform(c)}.
\end{remark}

\subsection{Choice of stabilization parameters and weight factor} \label{stabltermweightfactor}
 
\noindent The stabilization parameters $\beta_{\tau}, \beta_{n}$ and the weights $\gamma^{(1)}, \gamma^{(2)}$ play a key role in the numerical performance of a Nitsche-based stabilized FEM~\citep{annavarapu2012robust}. For instance, taking a small value for the stabilization parameters may undermine the positive definiteness of the global linear system of equations; whereas, taking a large value for the stabilization parameters will essentially lead to a penalty-like method for enforcing the interface constraints \citep{sanders2012nitsche}. Moreover, taking equal weights for dissimilar material interfaces may hamper numerical performance, or may not alleviate traction field oscillations. For constant strain triangular and tetrahedral elements, Annavarapu et al. \citep{annavarapu2012robust} provided estimates for the stabilization parameters using a local coercivity analysis as given by 
\begin{equation}
\beta_{n}=\beta_{\tau} =2~\Bigg(\frac{|\mathbf{D}^{(1)}|(\gamma^{(1)} )^{2}}{\textrm{meas}(\Omega^{(1)})}+\frac{|\mathbf{D}^{(2)}|(\gamma^{(2)} )^{2}}{\textrm{meas}(\Omega^{(2)})}\Bigg)~\textrm{meas}(\Gamma _{*}).
\label{eq:StabilizationParameters}
\end{equation}
Here, we simply use the above estimate of stabilization parameters for bilinear quadrilateral finite elements and conduct parametric sensitivity studies to illustrate their adequacy. Equation  \eref{eq:StabilizationParameters} establishes a functional dependency between the stabilization parameter and interface weights, and the weights are chosen so that it minimizes the stabilization parameter while ensuring coercivity. Here, we simply use expression for interface weights derived in \citep{annavarapu2012robust}:
\begin{equation}
\gamma^{(1)} =\dfrac{\dfrac{\textrm{meas}(\Omega^{(1)})}{|\mathbf{D}^{(1)}|}}{\dfrac{\textrm{meas}(\Omega^{(1)})}{|\mathbf{D}^{(1)}|}+\dfrac{\textrm{meas}(\Omega^{(2)})}{|\mathbf{D}^{(2)}|}}; \gamma^{(2)} = 1 - \gamma^{(1)},
\label{eq:weights}
\end{equation}
where $|\mathbf{D}|$ denotes the two-norm of the elasticity tensor, $\textrm{meas}(\Omega)$ denotes the area of neighboring bulk element in 2D, and $\textrm{meas}(\Gamma _{*})$ is the length of the interface element. 

\begin{remark}
For the weak form in \eqref{eq:stabilizedweakform}, precise estimates for the stabilization parameter can be derived for constant strain elements following the procedure described in \citep{annavarapu2012robust}. For higher-order elements, closed-form analytical estimates for the stabilization parameters are yet to be derived. However, the stabilization parameters can be specified by solving a local eigenvalue problem \citep{embar2010imposing}. 
\end{remark}

\section{Numerical Implementation and Model Parameters}\label{sec:NumImp}

We implemented the stabilized FEM in the commercial software ABAQUS through user-defined subroutines. In this section, we briefly present key details of ABAQUS implementation and list the model parameters that are specific to delamination analysis. The full details of the numerical implementation (omitted here), including the finite element approximation, discretization and linearization of the standard and stabilized weak forms, and the expressions for continuum and interface element force vectors and matrices can be found in \citep{ghosh2019stabilized}. 

\subsection{ABAQUS implementation}\label{sec:ABAQUSimplement}

We use bilinear quadrilateral four-noded plane stress continuum elements with four-point Gauss integration scheme and four-noded linear zero-thickness interface elements with two-point Gauss integration scheme. Although our method can be implemented in 3-D, we use the 2D plane stress approximation as it has been extensively used in prior studies and has been validated with experimental data \citep{harper2008cohesive,turon2010accurate,park2009unified,Versino2015}. We chose the finite element mesh and interface element size so that there are at least three interface elements within the estimated cohesive process zone; this is necessary for an accurate representation of the numerical stress distribution within the process zone at the point of initial crack propagation, as elaborated in  \citep{harper2008cohesive}. User-element subroutines in ABAQUS typically require the user to provide the stiffness matrix (AMATRX) and the right hand side (RHS) force vector. In our implementation, we utilize the UELMAT subroutine to provide the continuum element force vector and stiffness matrix, and the UEL subroutine to provide the cohesive element force vector and stiffness matrix. The UELMAT subroutine allows the user to access some of the inbuilt material models through utility subroutines \texttt{MATERIAL\char`_LIB\char`_MECH}, unlike the UMAT subroutine. Using global modules, we store and share the stress and shape function derivative matrices calculated in the UELMAT subroutine to the UEL subroutine for computing interface force vector and stiffness matrix. A detailed description of these computations can be found in \citep{ghosh2019stabilized} for isotropic elasticity; whereas, here we use anisotropic elasticity for composites. For the sake of verification or comparison in some simulation studies, we used ABAQUS in-built 4-noded 2D cohesive elements (COH2D4) along with 4-noded plane stress 2D continuum elements (CPS4). 

\subsection{Secant stiffness and convergence}
The anisotropic bilinear cohesive law described in Section \ref{IntrinsicCohesiveLaw} is highly nonlinear, because the scalar damage variable is a complex nonlinear function of the normal and tangential separations. The consistent tangent stiffness corresponding to this cohesive law can be derived from \eref{eq:BilinearTSLaw} as
\begin{equation}
\mathbf{K}^{\textrm{tan}} = -
\begin{bmatrix}
\pderiv{t_{\tau}}{\delta_{\tau}} & \pderiv{t_{\tau}}{\delta_{n}}\\ 
\pderiv{t_{n}}{\delta_{\tau}} & \pderiv{t_{n}}{\delta_{n}}
\end{bmatrix} = 
\begin{bmatrix}
(1-d_{s}) \alpha_{\tau}^0 - \alpha_{\tau}^0 \delta_{\tau} \pderiv{d_s}{\delta_{\tau}} & - \alpha_{\tau}^0 \delta_{\tau} \pderiv{d_s}{\delta_{n}}\\
- \alpha_{n}^0 \delta_{n} \dfrac{\langle \delta_{n} \rangle}{\delta_{n}} \pderiv{d_s}{\delta_{\tau}} & \left(1-d_{s} \dfrac{\langle \delta_{n} \rangle}{\delta_{n}}\right) \alpha_n^0 - \alpha_{n}^0 \delta_{n} \dfrac{\langle \delta_{n} \rangle}{\delta_{n}} \pderiv{d_s}{\delta_{n}}
\end{bmatrix}, \label{eq:ConsistentTangent}
\end{equation}
Deriving and implementing a closed-form expression of the above consistent tangent is arduous, and using it can cause numerical issues, as the diagonal terms become negative in the softening portion of the cohesive law. In our previous work \citep{liu2012discrete,jimenez2014discrete,jimenez2016parametric}, we argued that the secant stiffness is more advantageous with this cohesive law and demonstrated its accuracy and convergence with an  implicit scheme. The simpler secant stiffness matrix can be derived from \eref{eq:BilinearTSLaw} as
\begin{equation}
\mathbf{K}^{\textrm{sec}} = 
\begin{bmatrix}
(1-d_{s}) \alpha_{\tau}^0 & 0\\
0 & \left(1-d_{s} \dfrac{\langle \delta_{n} \rangle}{\delta_{n}}\right) \alpha_n^0
\end{bmatrix}, \label{eq:SecantTangent}
\end{equation}

Although we include the simple, linearized secant stiffness terms in the AMATRX defined in the UEL subroutine, the nonlinearity of the cohesive fracture problem is handled in ABAQUS/Stan-dard outside of the user subroutines. As detailed in the user manual \citep[][Chapter 7: Analysis Solution and Control]{ABAQUSManual}, ABAQUS/Standard combines incremental and iterative (Newton-Raphson) procedures for solving nonlinear problems. Using the secant stiffness necessitates smaller load/disp-lacement increments (pseudo-time steps) to gradually reach the final applied load/displacement. The user typically suggests the maximum and minimum increment size and the size of the first increment, and ABAQUS/Standard automatically chooses the size of the subsequent increments. Within each increment, ABAQUS/Standard automatically performs iteration to find an equilibrium solution based on a user-defined criteria for residual force and displacement correction. In all of our simulations, we use sufficiently small increment size that most displacement steps converge in a single iteration, except at certain time steps where more than one cohesive element fails. Also, in cohesive fracture simulations, the ABAQUS/Standard default criteria for residual force tolerance may be too small that numerical convergence may not be attainable as cohesive elements fail. Especially, using the secant stiffness it becomes necessary to increase these tolerances appropriately (sometimes by two orders of magnitude than the default value) to attain convergence. Therefore, we compare the predicted load--displacement curves against analytical solutions, experimental data, other numerical model results to ensure the accuracy of our simulations.

\subsection{Bulk material Properties}
We consider HTA/6376C unidirectional carbon-fiber-reinforced epoxy laminate as the generic composite material. Here, we only consider delamination or debonding along 2D straight interfaces between two HTA/6376C lamina with the fibers aligned either in the $X_1$ (in-plane horizontal) or $X_3$ (out-of-plane) direction, as indicated in Figure \ref{fig:domain}. According to the standard notation used to define stacking sequences in multi-ply composites laminates \citep{reddy2018practical}, ``[0/0]" laminate denotes the two-ply specimen with fibers in the top and bottom lamina oriented in $X_1$ direction. Similarly, ``[0/90]" laminate denotes the cross-ply specimen with the fibers in the top lamina oriented in the $X_1$ direction and fibers in the bottom lamina oriented in the $X_3$ direction. Usually, this notation implies that each laminate layer has the same thickness and made of the same composite material (HTA/6376C in our study). Unfortunately, experimental data from delamination tests is only available for [0/0] laminates. The anisotropic (transversely isotropic) linear elastic material properties for the 0$^\circ$ ply HTA/6376C laminate are listed in Table \ref{table:materialpropertiesDCB}, which is directly obtained from experiments \citep{asp2001delamination}. Using coordinate transformation relations we can easily obtain these material properties for the 90$^\circ$ ply HTA/6376C laminate.  

\begin{table}[ht]
\centering
\caption{Material properties of carbon fiber/epoxy laminated composite HTA/6376C obtained from \citep{asp2001delamination}}.
\begin{tabular}{cccccc}
\hline
  E$_{11}$  &          E$_{22}= E_{33}$ & G$_{12}=G_{13}$ & G$_{23}$ & $\nu_{12}= \nu_{13}$ & $\nu_{23}$   \\
                            (N/mm$^{2}$) &   (N/mm$^{2}$)        &  (N/mm$^{2}$) & (N/mm$^{2}$) & &  \\ \hline
   $ 1.2 \times 10^{5}$ &       $1.05 \times 10^{4}$ &       $5.52 \times 10^{3}$ & $3.48 \times 10^{3}$  & 0.3 & 0.51        \\
\label{table:materialpropertiesDCB}
\end{tabular}
\end{table} 

\subsection{Cohesive zone model parameters}

The CZM parameters chosen in our simulation studies are listed in Table \ref{table:cohesivepropertiesDCB}. The mode I and mode II fracture energies for the 0$^\circ$ ply HTA/6376C laminate are taken from \citep{asp2001delamination}. For ensuring the accuracy and convergence of delamination analysis using the FEM with cohesive elements, two conditions must be satisfied \citep{Turon2007}: (1) the element size must be less than the cohesive (process) zone length, which is determined by fracture toughness and cohesive strength; and (2) the cohesive stiffness must be large enough to avoid the introduction of artificial compliance.  
The cohesive strength is often chosen based on cohesive zone length and mesh size considerations \citep[see Eq. (7) in Ref.][]{harper2008cohesive}, owing to computational cost or limitations. Choosing a small cohesive strength may yield a poor peak load prediction, but beyond a certain value choosing a larger cohesive strength will not improve model fit with the load--displacement data from quasi-static delamination tests, but will severely restrict the mesh size and increase computational cost. Here, we use the cohesive strength values suggested in \citep{harper2008cohesive} to ensure numerical accuracy and efficiency. Due to the unavailability of experimental data for the 90$^\circ$ ply HTA/6376C laminate, we assume the same CZM parameters listed in Table \ref{table:cohesivepropertiesDCB}. 

The cohesive stiffness is generally considered to be a penalty parameter and various guidelines have been proposed in the literature for selecting the stiffness. Although the purpose of the cohesive stiffness in intrinsic/extrinsic cohesive zone models is to account for the elastic loading, unloading and reloading response of the fracture/delamination interface, it can contribute to the global deformation response and introduce artificial compliance or numerical instability issues. Based on 1D laminate model, the cohesive stiffness necessary to avoid artificial compliance issue can be estimated as \citep{Turon2007}
\begin{equation}
    \alpha^0 = E~M/t, \label{eq:StiffnessEstimate}
\end{equation}
where $E$ is the Young's modulus of the material along the laminate thickness direction, $t$ is the sub-laminate thickness , and $M$ is a non-dimensional number that is to be chosen much larger than one. For cohesive interfaces in non-laminates $t$ is not defined, so in \eref{eq:StiffnessEstimate} it can be replaced by a certain length measure $h$ of the bulk material \citep{song2006bilinear} or the finite element mesh size. Taking $M=100$, $E=E_{22}=1.05\times 10^4$ N/mm$^2$ and $t=1.55$ mm for the delamination tests (see Section \ref{sec:NumEx}), we estimate $\alpha^0\approx 10^6$ N/mm$^3$. To demonstrate the performance of the standard and stabilized methods, we assume three values of cohesive stiffness in our studies, including values that are two orders of magnitude smaller and larger than the above estimate. However, we note that for thin-ply laminates with $t<0.1$ mm or for small values of length measure $h<0.1$ mm in non-laminates, the estimated cohesive stiffness $\alpha^0> 10^7$ N/mm$^3$.

\begin{table}[ht]
\centering
\caption{Cohesive zone model parameters for the carbon fiber/epoxy laminated composite HTA/6376C are taken from \citep{harper2008cohesive}, except the cohesive stiffness values.}
\begin{tabular}{cccccc}
\hline
  $\alpha_n^0$  &          $\alpha_{\tau}^0$ & G$_{IC}$  & G$_{IIC}$  & $\sigma_{\text{max}}$  & $\tau_{\text{max}}$  \\
                            (N/mm$^{3}$) &   (N/mm$^{3}$)        &  (N/mm) &  (N/mm) &  (N/mm$^{2}$) &  (N/mm$^{2}$) \\ \hline
   $  \{10^4, 10^6, 10^8\} $ &       $\{10^4, 10^6, 10^8\}$ &       0.26 &          1.002 &        30 &          60 \\
\label{table:cohesivepropertiesDCB}
\end{tabular}
\end{table} 

\section{Numerical Examples}\label{sec:NumEx}

In this section, we present several examples to demonstrate the ability of the Nitsche-inspired stabilized formulation in alleviating oscillations in interface traction using constant strain patch tests, and pure mode I, mode II and mixed mode delamination tests. Through these tests, we specifically examine numerical stability at similar and dissimilar laminate interfaces defined by anisotropic and isotropic cohesive laws using perturbed, structured and unstructured meshes.

\subsection{Patch Tests}\label{Patch Test}
We assess the ability of the standard and stabilized formulations in alleviating traction oscillations at horizontal and inclined straight interfaces using the constant strain patch test. Under compressive loading, we assume a stiff elastic response in the normal direction to enforce contact and a weak elastic sliding response in the tangential direction, which is captured by the anisotropic CZM. We assign the square plate with a side length $L = 1$ mm and the horizontal delamination interface at mid-height. To apply the compressive load, we constrain both vertical and horizontal displacements at the bottom edge of the plate, and prescribe a uniform vertical displacement $\Delta= -0.1$ mm at the top edge of the plate. We specify traction-free conditions at the left and right edges of the square plate. The cohesive parameters and material properties assumed for this test are listed in Tables \ref{table:materialpropertiesDCB} and \ref{table:cohesivepropertiesDCB}, respectively. 

\subsubsection{Square plate with horizontal  interface}\label{Perturbed Interface}

\begin{figure}[ht]
  \centering
 \includegraphics[width=0.7\textwidth]{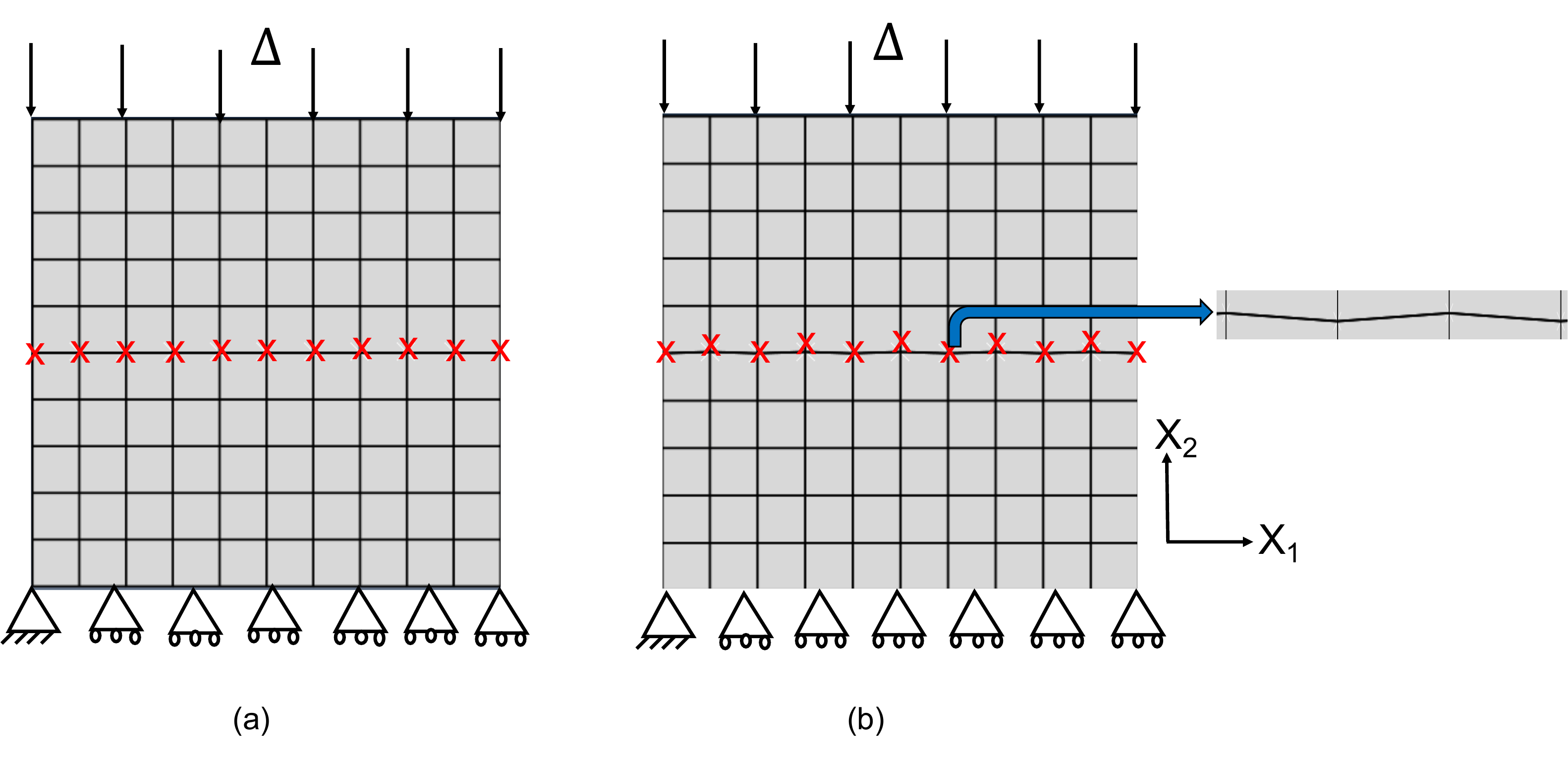}
 \caption{Boundary conditions and mesh used for the square plate with horizontal interface: (a) straight interface; (b) perturbed interface. The nodes are perturbed by $\approx$ 3$\%$ of the element size and a zoom of the interface undulations is shown in the inset.}
 \label{fig:unitsquarehorzcrack}
\end{figure}

\begin{figure}[ht]
  \centering
 \includegraphics[width=0.7\textwidth]{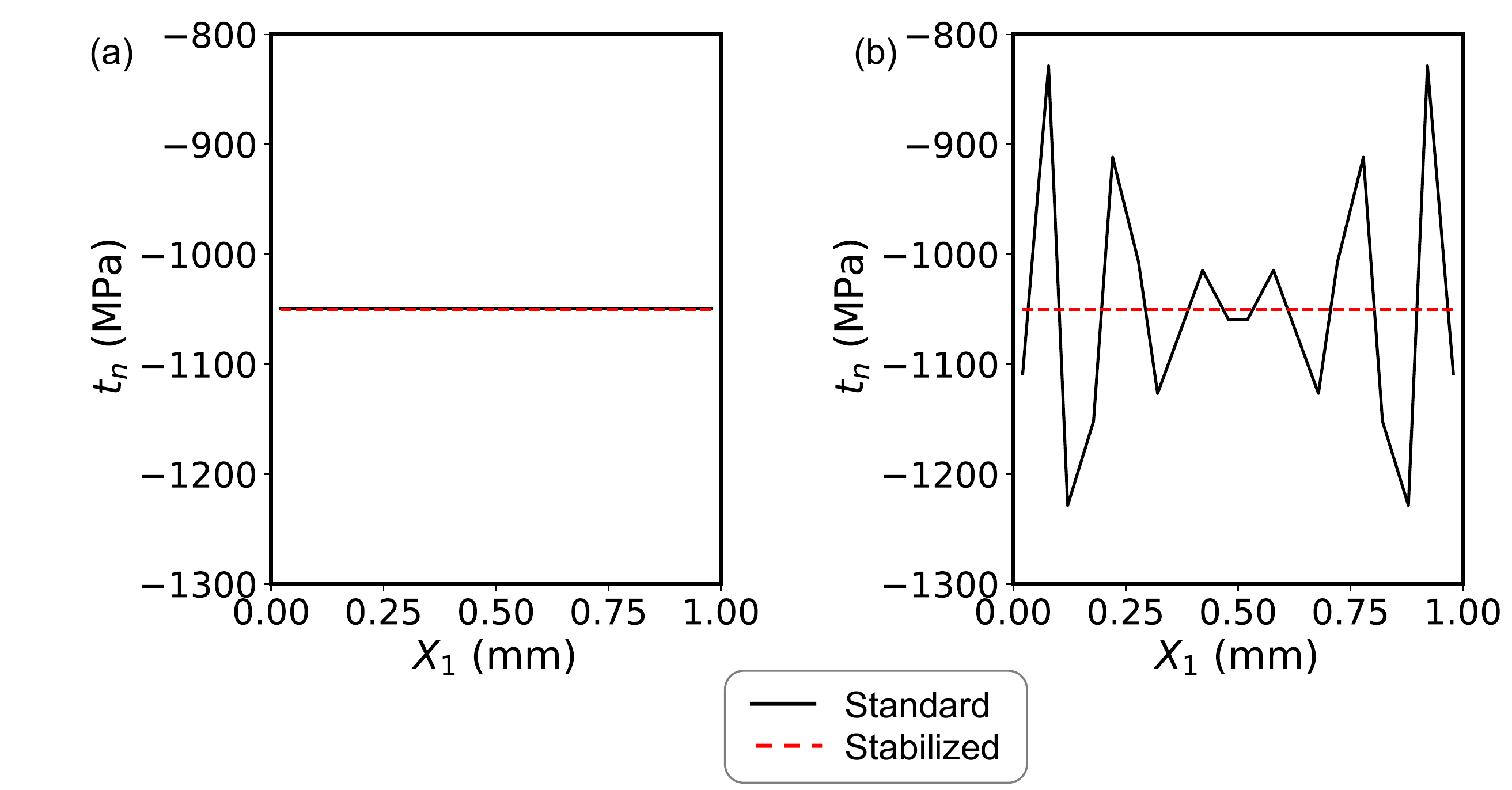}
 \caption{Traction profiles obtained from the standard and stabilized methods with the anisotropic CZM ($\alpha_n^0= 10^{8}$ and $\alpha_{\tau}^0= 10^{1} $ N/mm$^{3}$) for the square plate made of [0/0] laminate: (a) horizontal interface (b) perturbed interface.} \label{fig:horzinterfacetraction}
\end{figure}

To examine mesh sensitivity, we generate a $10 \times 10$ structured square mesh with element length of 0.1 mm (Fig. \ref{fig:unitsquarehorzcrack}a) and perturb the interface nodes by $\approx 3\%$ of the element length (Fig. \ref{fig:unitsquarehorzcrack}b). In Fig. \ref{fig:horzinterfacetraction}, we show the normal traction profile versus the horizontal coordinate along the [0/0] laminate interface obtained from the standard and stabilized methods. We consider high stiffness-contrast with the anisotropic CZM, where $\alpha^{0}_{n}$= 10$^{8}$ N/mm$^{3}$ and $\alpha^{0}_{\tau}$= 10$^{1}$ N/mm$^{3}$. According to \eref{eq:StabilizationParameters}, we take the stabilization parameters $\beta_n = \beta_{\tau} = 3 \times 10^{6}$  N/mm$^{3}$. As shown in Fig. \ref{fig:horzinterfacetraction}a, if the interface is perfectly flat, then both standard and stabilized methods yield smooth traction profiles without any spurious oscillations. However, we note that the standard FEM exhibits spurious traction oscillations even with the unperturbed mesh, if the normal stiffness $\alpha^{0}_{n} = \alpha^{0}_{\tau} \geq 10^{11}$ N/mm$^{3}$ (results not shown here). From Fig. \ref{fig:horzinterfacetraction}b, we can see that the standard FEM suffers from severe numerical instability even with a slightly perturbed interface mesh, as evident from the large amplitude traction oscillations. In contrast, our stabilized FEM is stable and alleviates traction oscillations for anisotropic CZMs with high stiffness contrast between normal and tangential directions. 

\subsubsection{Square plate with inclined interface}\label{Inclined Interface}

\begin{figure}[ht]
  \centering
 \includegraphics[width=0.5\textwidth]{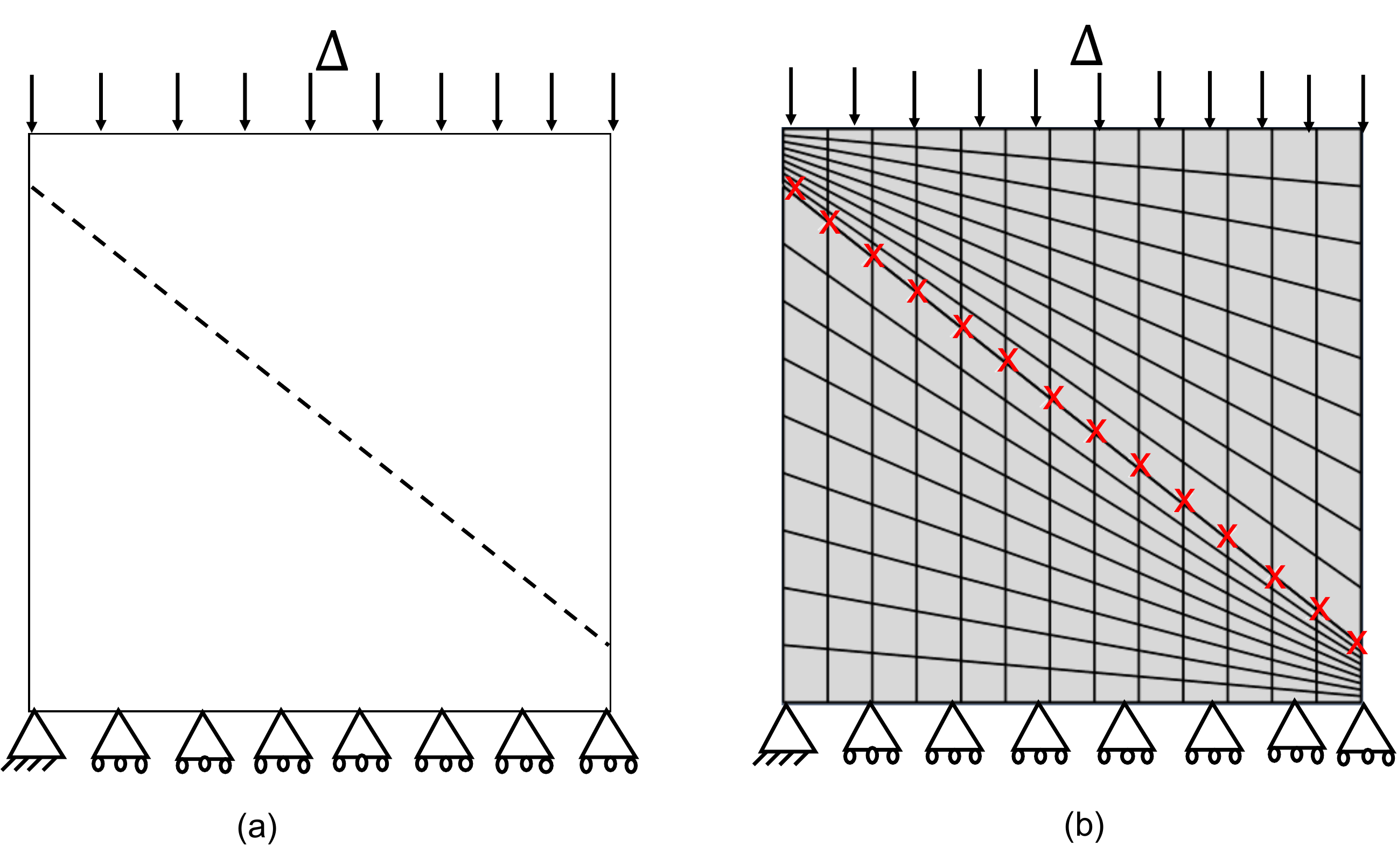}
 \caption{Boundary conditions and mesh used for the square plate with inclined interface: (a) schematic diagram; (b) structured finite element mesh with quadrilateral (non-rectangular) elements.}
 \label{fig:unitsquarediagcrack}
\end{figure}

\begin{figure}[ht]
  \centering
 \includegraphics[width=0.7\textwidth]{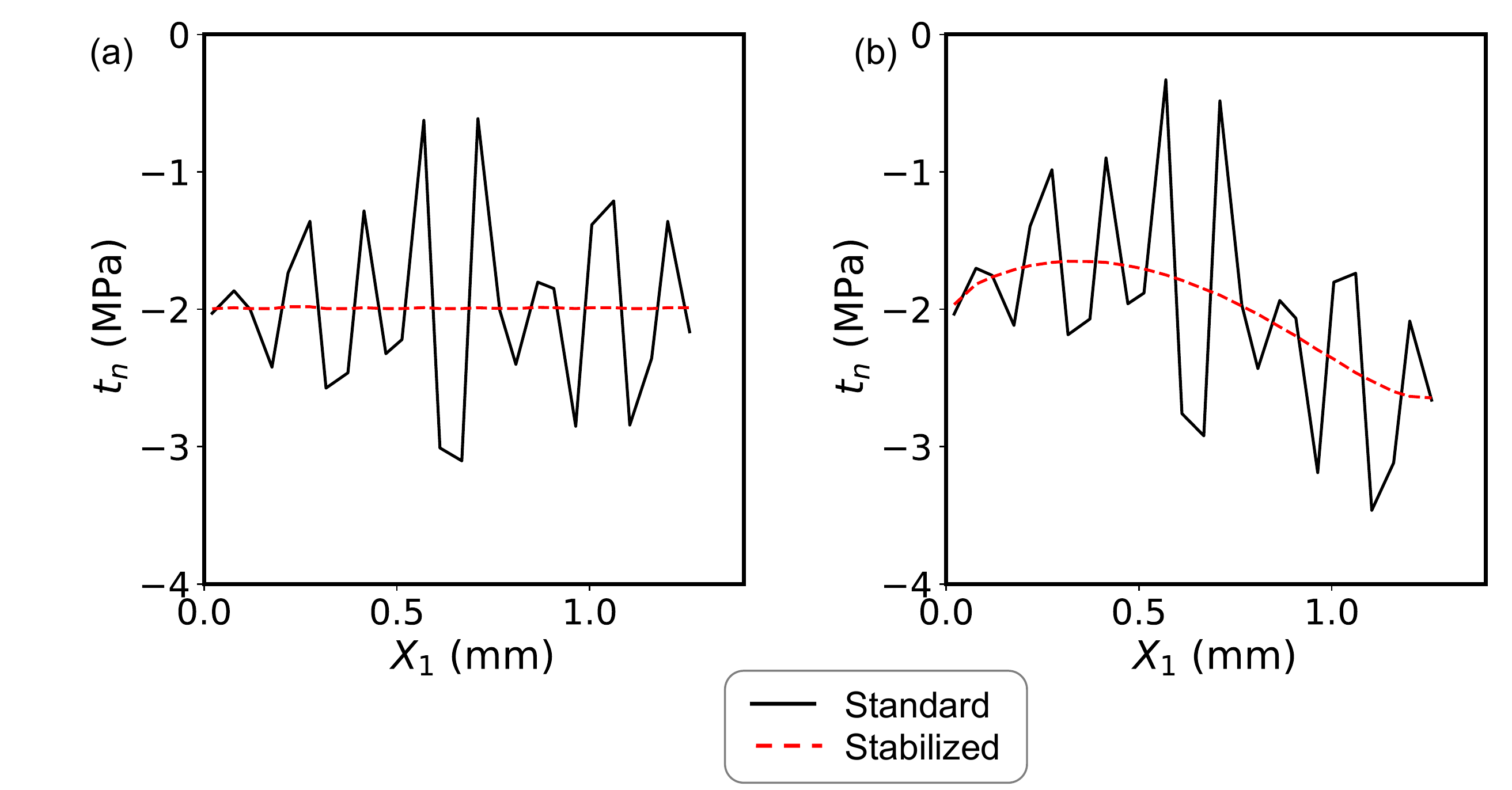}
 \caption{Traction profiles obtained from the standard and stabilized methods with the anisotropic CZM ($\alpha_n^0= 10^{8}$ and $\alpha_{\tau}^0= 10^{1} $ N/mm$^{3}$) for the square plate with an inclined interface: (a) [0/0] laminate (b) [0/90] laminate.} \label{fig:inclinedinterfacetraction}
\end{figure}

We consider a straight interface inclined at an initial angle of 140.4$^\circ$ with the horizontal $X_1$ axis embedded within the square plate of side length $L = 1$ mm (Fig. \ref{fig:unitsquarediagcrack}a). We discretize the domain using a $13 \times 18$ semi-structured mesh with quadrilateral element, so that the interface is divided into 13 elements (Fig. \ref{fig:unitsquarediagcrack}b). According to \eref{eq:StabilizationParameters}, we take the stabilization parameters $\beta_n = \beta_{\tau} = 5 \times 10^{6}, 3 \times 10^5$  N/mm$^{3}$ for [0/0] and [0/90] laminate interfaces, respectively. We consider the anisotropic CZM with high stiffness contrast, where $\alpha^{0}_{n}$= 10$^{8}$ N/mm$^{3}$ and $\alpha^{0}_{\tau}$= 10$^{1}$ N/mm$^{3}$. In Fig. \ref{fig:inclinedinterfacetraction}, we show the normal traction profile versus the horizontal $X_1$ coordinate of the integration points along the inclined interface for [0/0] and [0/90] laminates obtained from the standard and stabilized methods. While the standard FEM exhibits instability evident from the large amplitude traction oscillations, the stabilized FEM is able to alleviate oscillations and yields a smooth traction profile. This study illustrates the drawback of the standard formulation for semi-structured meshes and potentially unstructured meshes when using anisotropic CZMs with high stiffness contrast. Overall, the two patch tests highlight the robustness of our stabilized FEM, but in the following sections we will investigate its performance for benchmark delamination tests.

\subsection{Double Cantilever Beam Test}\label{Mode-I delamination test}
\begin{figure}[ht]
\centering
\includegraphics[width=1\textwidth]{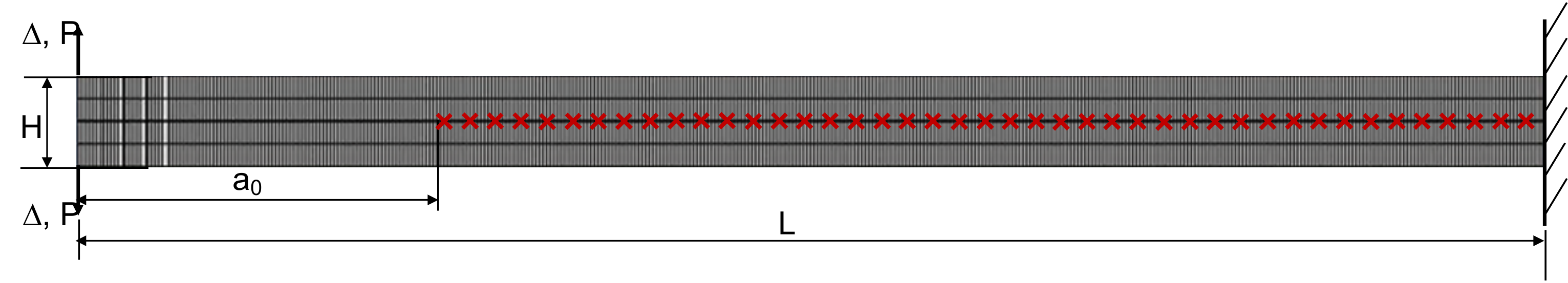}
\caption{Geometry and boundary conditions for the double cantilever beam test. The dimensions are: L = 150 mm, H = 3.1 mm and a$_{0}$ = 35 mm.}
\label{fig:DCBTestSetup}
\end{figure}

We investigate the ability of the stabilized FEM in recovering oscillation-free interface traction during mode-I delamination crack growth, using the double cantilever beam (DCB) test. The specimen geometry and test set-up along with the finite element mesh are shown in Fig. \ref{fig:DCBTestSetup}. To initiate the delamination process, a pre-crack is placed at the left end of the beam, and equal and opposite vertical displacements ($\Delta$) are applied on the upper and lower nodes. The corresponding load (P) is determined from the simulation using the reaction force at the corresponding node. The fixed boundary condition is applied at the right end of the beam. 

\begin{figure}[ht]
 \centering
 \begin{subfigure}[b]{0.4\linewidth}
   \centering\includegraphics[width=\textwidth]{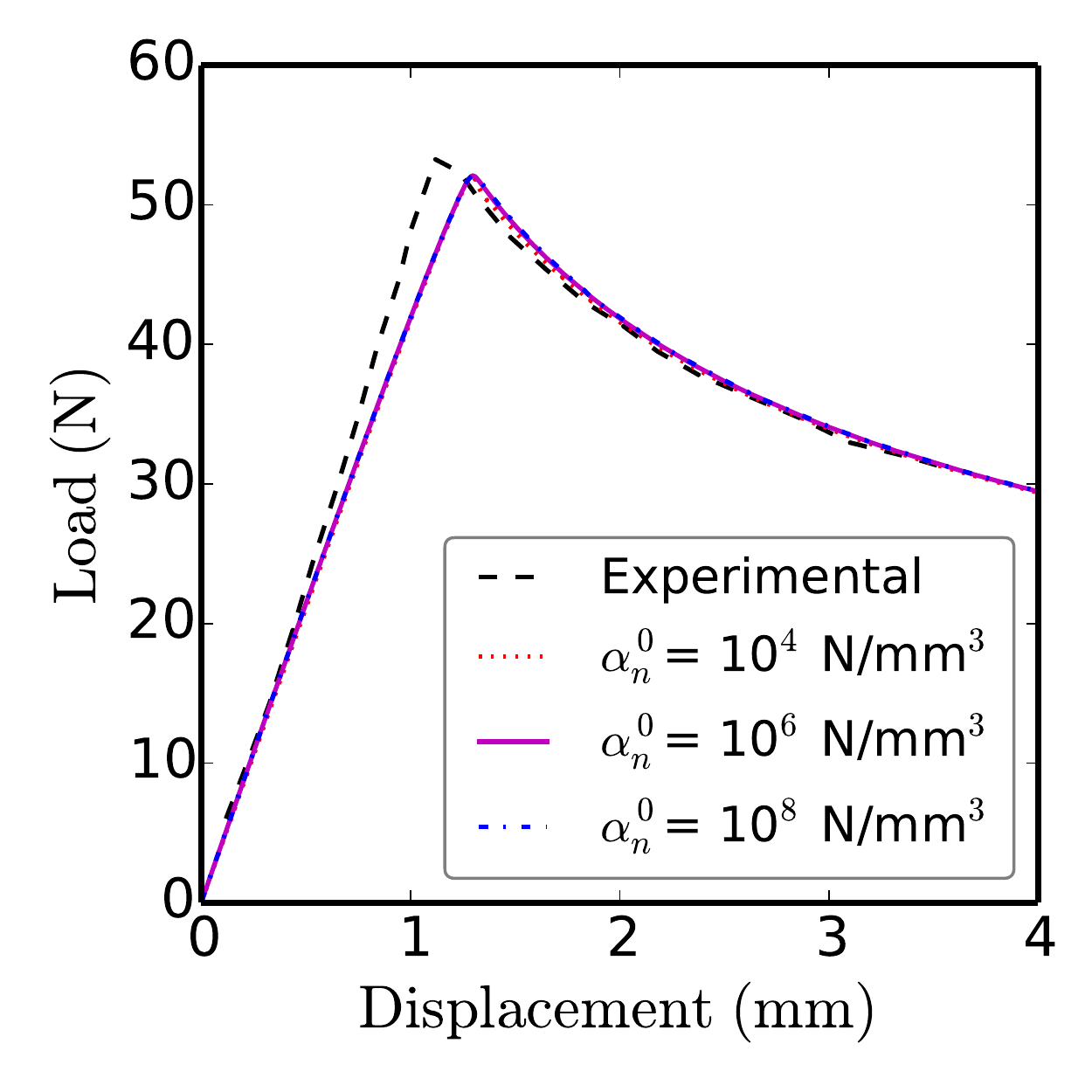}
   \caption{}
       \label{fig:stabilizedLdispDCB00}
 \end{subfigure}%
 ~
 \begin{subfigure}[b]{0.4\linewidth}
   \centering\includegraphics[width=\textwidth]{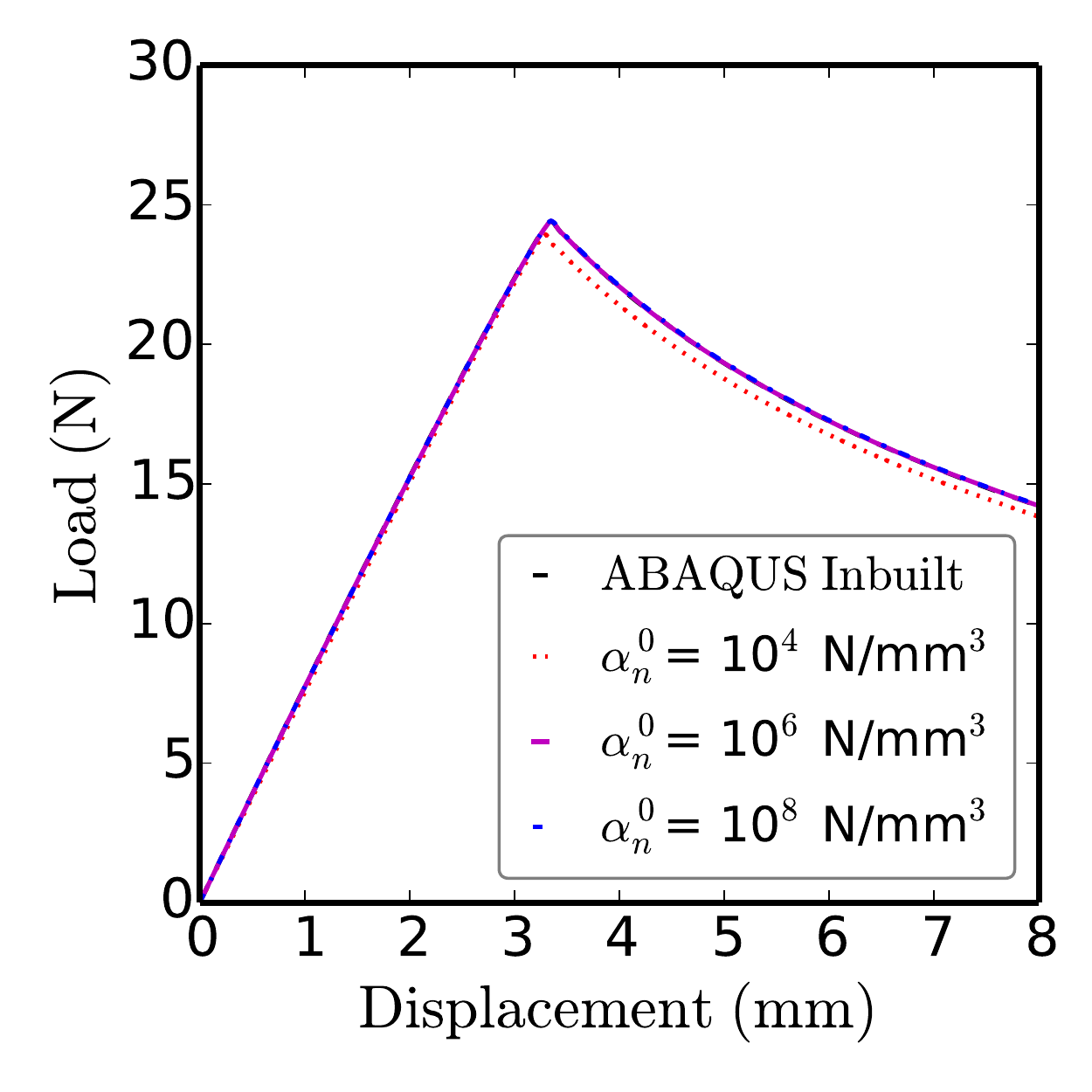}
   \caption{}
       \label{fig:stabilizedLdispDCB090}
 \end{subfigure}
 \caption{Load versus displacement curves for the double cantilever beam test obtained from the stabilized FEM: (a) [0/0] laminate and (b) [0/90] laminate. } \label{fig:DCBstabilizedloaddisp}
\end{figure}

In Fig. \ref{fig:stabilizedLdispDCB00}, we compare the load--displacement curves obtained from the stabilized formulation for [0/0] laminate with experimental data from \citep{hallett2009experimental}, to check that our choice of strength and displacement increment is appropriate and to determine the sensitivity to cohesive stiffness. Evidently, there is good agreement between the numerically predicted load-displacement response and the experimental data. The slight mismatch in the predicted peak load and the initial slope of the load--displacement curves is plausibly due to the idealization of boundary and loading conditions in the model, compared to the experimental test setup. In Fig. \ref{fig:stabilizedLdispDCB090}, we compare the load--displacement curves obtained from the stabilized FEM for [0/90] laminate with those obtained from ABAQUS inbuilt cohesive elements \citep{ABAQUSManual}, as experimental data is unavailable. The perfect match of load--displacement curves obtained with our user-defined and ABAQUS inbuilt (COH2D4) elements verifies the correctness of our implementation. The load--displacement curves obtained from the standard FEM exactly match with those from the stabilized FEM, so we do not show them here. As the two methods differ primarily in their ability to recover interface traction fields, we examine traction fields along the delamination interfaces obtained from standard and stabilized methods for [0/0] and [0/90] laminates. 

\subsubsection{Traction at [0/0] laminate interface}

\begin{figure}[ht]
\centering
\includegraphics[width=1\textwidth]{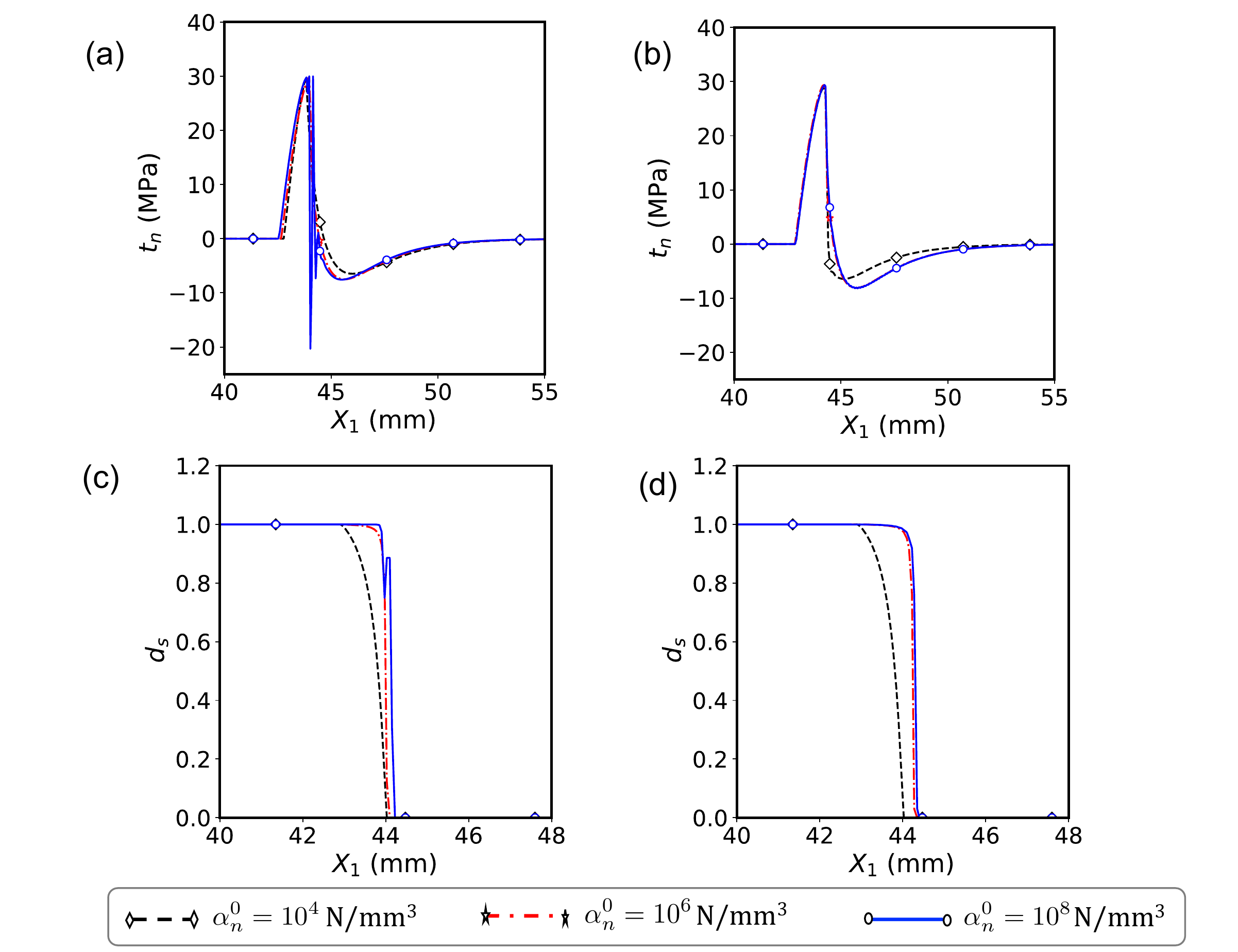}
\caption{Normal traction and damage versus interface length curves from the double cantilever beam test ([0/0] ply orientation) for different cohesive stiffness values: (a), (c) standard FEM and (b), (d) stabilized FEM.} \label{fig:DCB00}
\end{figure}

From Fig. \ref{fig:DCB00}a, we find that for smaller values of stiffness $\alpha_n^0= 10^{4}, 10^{6} $ N/mm$^{3}$ the normal traction profile from the standard FEM is smooth, but for the larger value of $\alpha_n^0= 10^{8}$ N/mm$^{3}$ significant oscillations are observed in the traction profile. From Fig. \ref{fig:DCB00}b, it is evident that the stabilized FEM is able to alleviate traction oscillations for the large value of cohesive stiffness. We also notice that the location of the region of traction oscillations (i.e. $X_1 \approx 44$ mm) coincides with the transition from tensile to compressive normal traction. To further examine the effect of this instability on interface damage, we plotted the respective damage profiles obtained from both methods in Figs. \ref{fig:DCB00}c and \ref{fig:DCB00}d. For $\alpha_n^0= 10^{8}$ N/mm$^{3}$ at the location of traction oscillations, we also find an oscillation in the damage profile with the standard FEM; whereas, there is no such oscillation in the damage profile with the stabilized FEM. This study shows evidence of instability with the standard FEM even with a perfectly flat interface and structured rectangular mesh.

\begin{figure}[ht]
\centering
\includegraphics[width=1\textwidth]{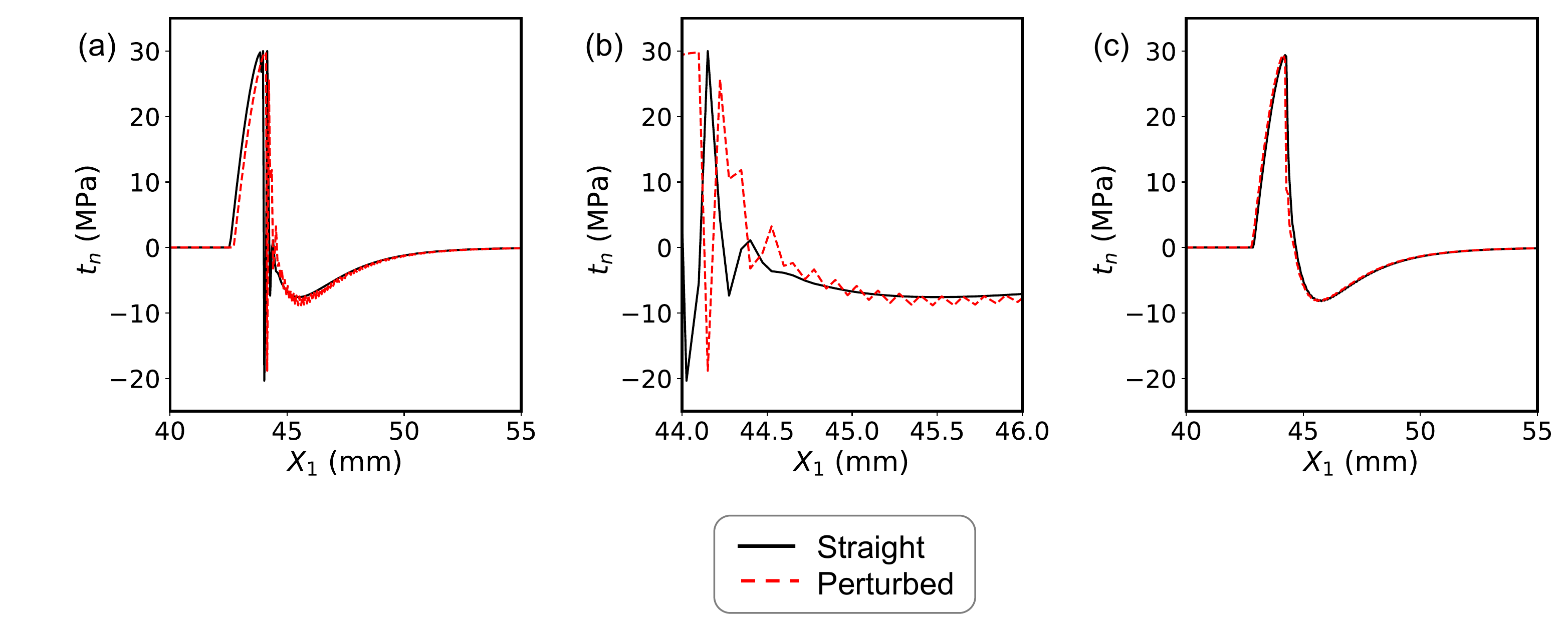}
\caption{Normal traction versus interface length for $\alpha_n^0= 10^{8}$ N/mm$^{3}$ obtained from DCB test ([0/0] ply orientation): (a) standard FEM, (b) zoom in of standard FEM near the crack tip and (c) stabilized FEM.} \label{fig:DCBPerturbed00}
\end{figure}

We next examine the interface traction along the delamination interface recovered with a perturbed rectangular mesh, where we change the coordinates of interface nodes by $\approx 3\%$ of the element length. In Fig. \ref{fig:DCBPerturbed00}a and \ref{fig:DCBPerturbed00}b, we show the normal traction profiles obtained from the standard FEM and a zoomed-in image showing evidence of spurious oscillations in the compression region of the traction. From Fig. \ref{fig:DCBPerturbed00}c it is evident that the stabilized FEM is able to alleviate the traction oscillations. This study demonstrates that numerical instability with the standard FEM is more pronounced under contact conditions (i.e. in the regions where the normal traction is negative) when using perturbed or unstructured finite element meshes, although the amplitude of the spurious oscillations in the mode I DCB test is generally small.  

\subsubsection{Traction at [0/90] laminate interface}

\begin{figure}[ht]
\centering
\includegraphics[width=1\textwidth]{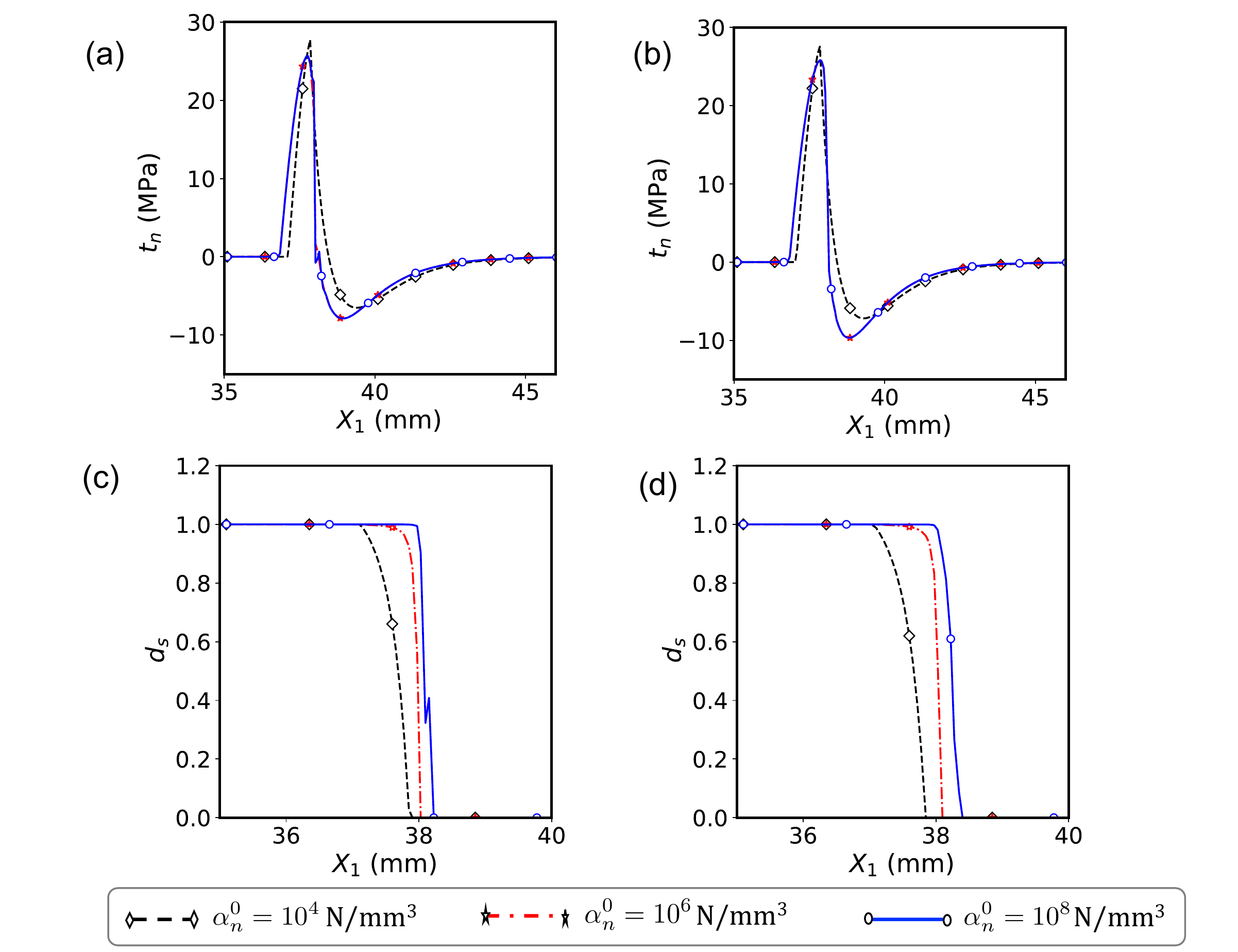}
\caption{Normal traction and damage versus interface length curves from the double cantilever beam test ([0/90] ply orientation) for different cohesive stiffness values: (a), (c) standard FEM and (b), (d) stabilized FEM.} \label{fig:DCB090}
\end{figure}

We now compare the interface traction fields recovered from standard and stabilized methods at dissimilar material interfaces. Recall that the 0$^\circ$ laminate exhibits anisotropic material behavior in the X$_1$ and X$_2$ plane; whereas, the 90$^\circ$ laminate exhibits isotropic material behavior in this plane with an order of magnitude contrast in elastic modulus ($E_{11}/E_{22}=11.4$). In Fig. \ref{fig:DCB090}, we show the normal traction along the delamination interface recovered from the standard and stabilized methods for different values of initial cohesive stiffness. We do find a minor oscillation in normal traction and damage profiles in Figs. \ref{fig:DCB090}a and \ref{fig:DCB090}c with the standard FEM for $\alpha_n^0= 10^{8}$ N/mm$^{3}$, which is not the case with the stabilized FEM. Notably, there are minor differences in the traction and damage curves for different stiffness values, unless stiffness is taken greater than $10^8$ N/mm$^3$. 

\subsection{End notch flexure test}\label{Mode-II delamination test}

\begin{figure}[ht]
\centering
\includegraphics[width=1\textwidth]{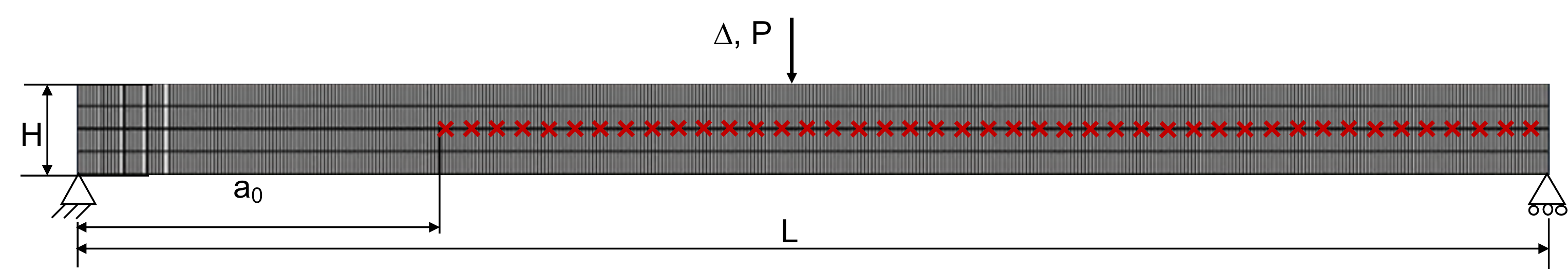}
\caption{Geometry and boundary conditions for the end notch flexure test. The dimensions are: L = 100 mm, H = 3.1 mm and a$_{0}$ = 35 mm.}
\label{fig:ENFTestSetup}
\end{figure}

\begin{figure}[ht]
 \centering
 \begin{subfigure}[b]{0.4\linewidth}
   \centering\includegraphics[width=\textwidth]{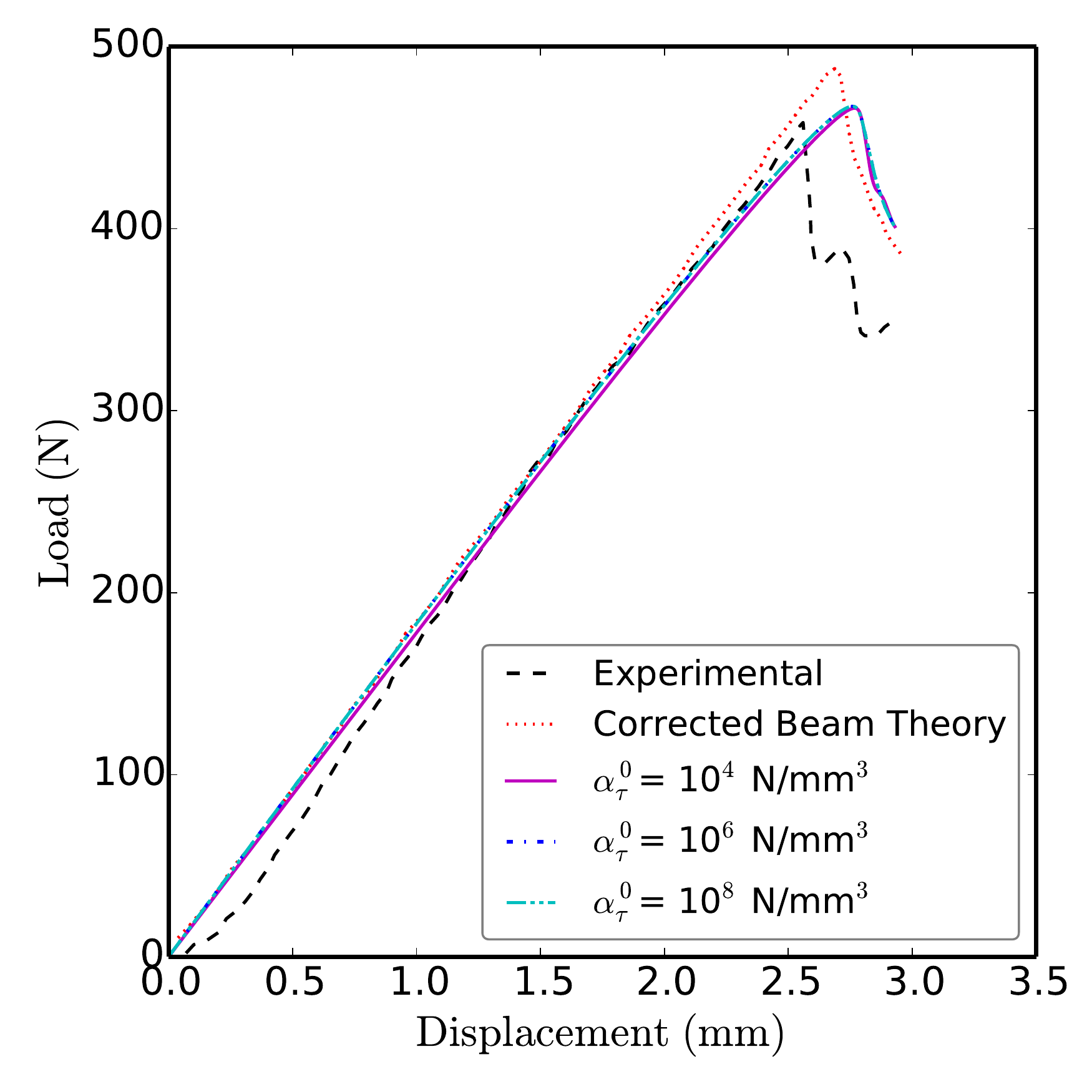}
   \caption{}
       \label{fig:standardENF00loaddisp}
 \end{subfigure}%
 ~
 \begin{subfigure}[b]{0.4\linewidth}
   \centering\includegraphics[width=\textwidth]{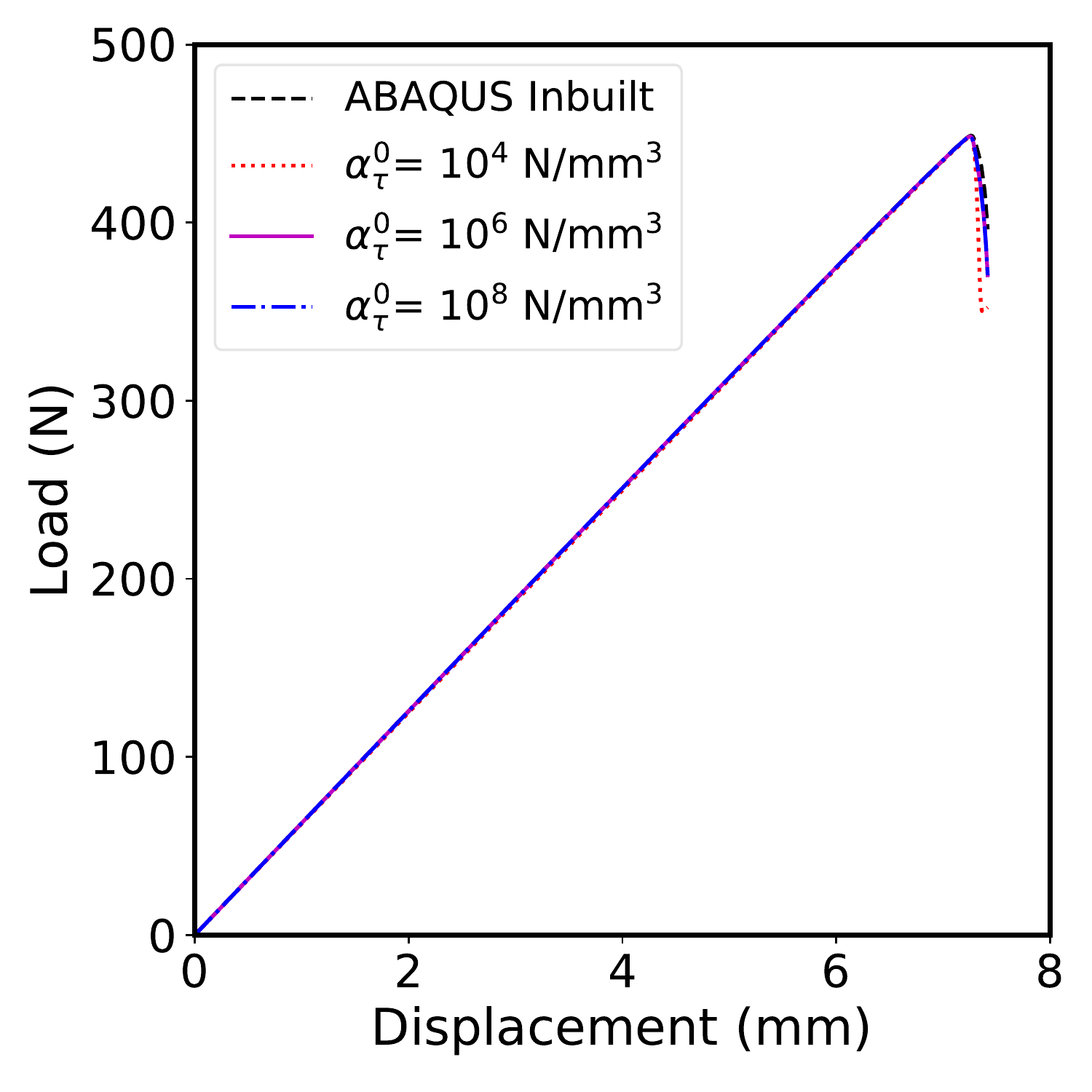}
   \caption{}
       \label{fig:stabilizedENF090loaddisp}
 \end{subfigure}
  \caption{Load versus displacement curves for the end notch flexure test obtained using the stabilized FEM: (a) [0/0] laminate and (b) [0/90] laminate.} \label{fig:ENFstablizedloaddisp}
\end{figure}

We investigate the ability of the stabilized FEM in recovering oscillation-free interface traction during mode-II delamination crack growth, using the end notch flexure (ENF) test. The specimen geometry and test set-up are shown along with the finite element mesh in Fig. \ref{fig:ENFTestSetup}. To initiate the delamination process, a pre-crack is placed at the left end of the beam and downward vertical displacement is applied at the middle of the simply supported beam. 
In Fig. \ref{fig:standardENF00loaddisp}, we compare the load--displacement curves obtained from the stabilized FEM for [0/0] laminate specimen along with the experimental data and the corrected beam theory solution from \citep{harper2008cohesive} for different values of cohesive stiffness. In Fig. \ref{fig:stabilizedENF090loaddisp}, we compare the load--displacement curves obtained from the stabilized FEM for [0/90] laminate with those obtained from ABAQUS inbuilt cohesive elements \citep{ABAQUSManual}, due to the unavailability of experimental data or analytical solutions. The good match of load--displacement curves, including the predicted peak load, obtained with our user-defined and ABAQUS inbuilt (COH2D4) elements verifies the correctness of our implementation. The load--displacement curves obtained from the standard FEM exactly match with those from the stabilized FEM, so we do not show them here. 
 
\subsubsection{Traction at [0/0] laminate interface}

\begin{figure}[ht]
\centering
\includegraphics[width=1\textwidth]{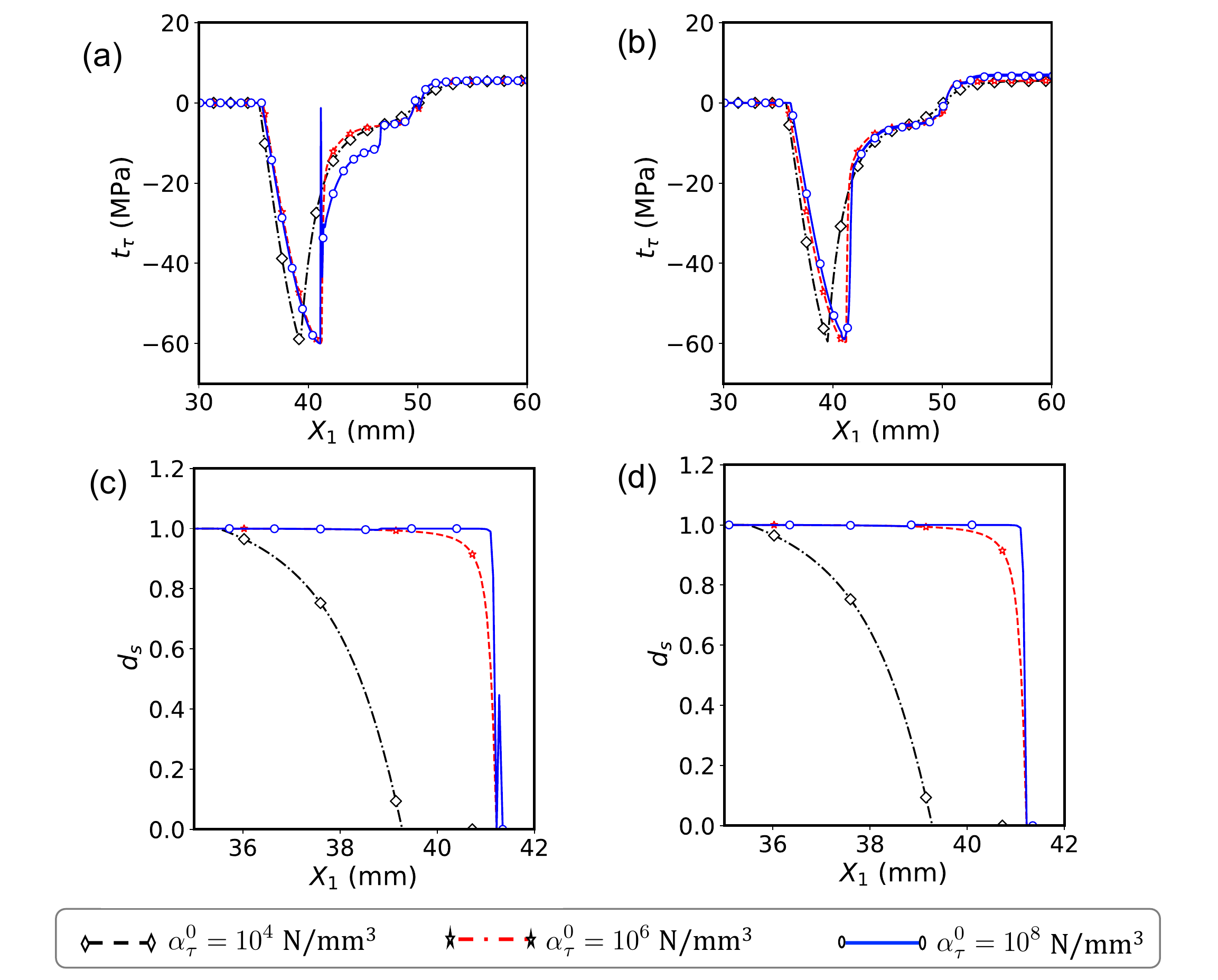}
\caption{Tangential traction and damage versus interface length curves from the end notch flexure test ([0/0] ply orientation) for different cohesive stiffness values: (a), (c) standard FEM and (b), (d)  stabilized FEM.} \label{fig:ENF00}
\end{figure}

From Fig. \ref{fig:ENF00}a, it is evident that the tangential traction profile obtained from the standard FEM is oscillation-free for smaller cohesive stiffness values $\alpha_{\tau}^0= 10^{4}, 10^{6} $ N/mm$^{3}$, but for the larger value $\alpha_{\tau}^0= 10^{8}$ N/mm$^{3}$ a significant oscillation can be seen. Also, the traction field for $\alpha_{\tau}^0= 10^{4}$ from the standard FEM does not match with that for larger stiffness values, which emphasizes the importance of choosing a large enough cohesive stiffness. The traction profile obtained from the stabilized FEM shown in Fig. \ref{fig:ENF00}b does not exhibit any oscillations for all cohesive stiffness values, thus demonstrating stability and robustness. We also plot the respective damage profiles obtained from both the methods in Figs. \ref{fig:ENF00}c and \ref{fig:ENF00}d, which show a corresponding oscillation in the damage profile for the standard FEM for $\alpha_{\tau}^0= 10^{8}$ N/mm$^{3}$, but no such oscillation exists in the damage profile with the stabilized FEM. These results indicate that even though the standard and stabilized methods capture the load displacement curves well, numerical instability can corrupt the tangential traction and damage field at the interlaminar interface. 

\begin{figure}[ht]
\centering
\includegraphics[width=1\textwidth]{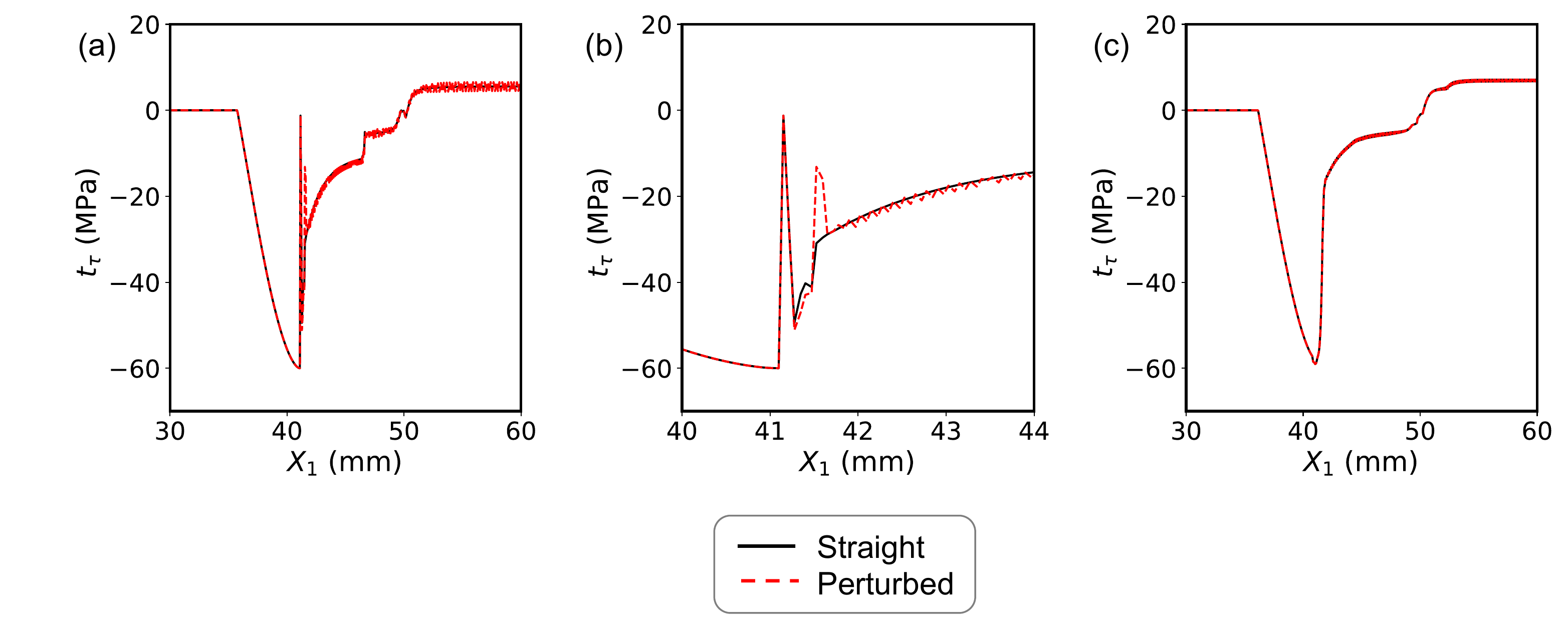}
\caption{Tangential traction versus interface length for $\alpha_{\tau}^0= 10^{8}$ N/mm$^{3}$ obtained from ENF test ([0/0] ply orientation): (a) standard FEM , (b) zoom in near the crack tip for standard FEM and (c) stabilized FEM.} \label{fig:ENFperturbed00}
\end{figure}

We next examined the tangential traction along the delamination interface recovered with a perturbed rectangular mesh, where we change the coordinates of interface nodes by $\approx 3\%$ of the element length. In Fig. \ref{fig:ENFperturbed00}a and \ref{fig:ENFperturbed00}b, we show the tangential traction profiles obtained from the standard FEM and a zoomed-in image showing clear evidence of spurious oscillations. From Fig. \ref{fig:ENFperturbed00}c it is evident that the stabilized FEM is able to alleviate the large amplitude oscillations. This study demonstrates that numerical instability with the standard FEM can corrupt tangential traction (i.e. shear stress at the interface) when using perturbed or unstructured finite element meshes, and the amplitude of some spurious oscillations in the mode II ENF test can be large (as much as 50\% local error). Therefore, we recommend using the stabilized FEM to improve accuracy and avoid convergence issues with stiff cohesive laws.
 
\subsubsection{Traction at [0/90] laminate interface} 
 
\begin{figure}[ht]
\centering
\includegraphics[width=1\textwidth]{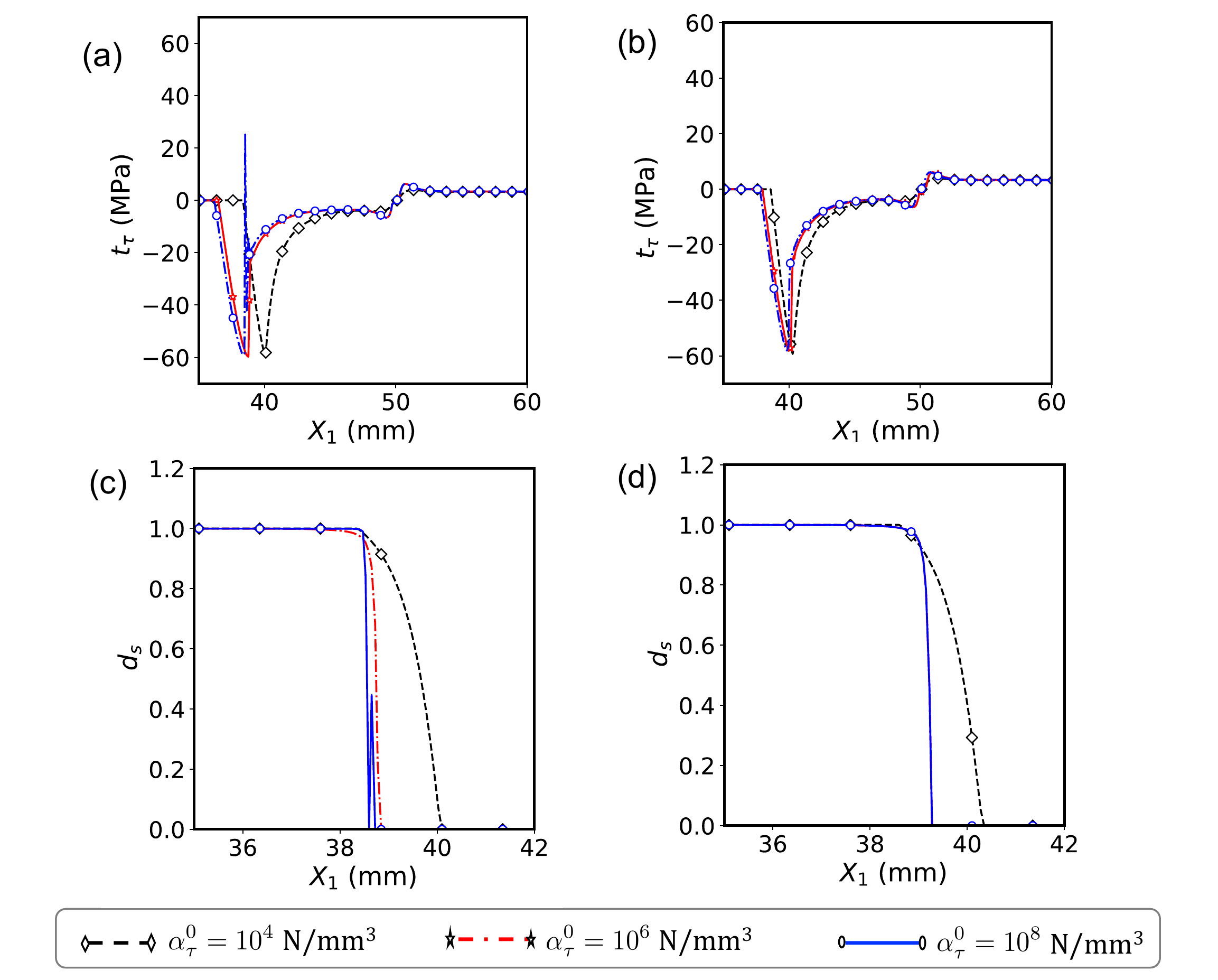}
\caption{ Tangential traction and damage versus interface length curves from the end notch flexure test ([0/90] ply orientation) for different cohesive stiffness values: (a), (c) standard FEM and (b), (d) stabilized FEM.} \label{fig:ENF090}
\end{figure}

We now examine the interface traction fields obtained from both standard and stabilized methods for the [0/90] laminate interface. Although we expect a similar response to the [0/0] laminate interface case, we present the results here for completeness. In Fig. \ref{fig:ENF090}, we show the tangential traction along the delamination interface recovered from the standard and stabilized methods for different values of initial cohesive stiffness. We do find large amplitude oscillations in the tangential traction for $\alpha_{\tau}^0= 10^{8}$ N/mm$^{3}$ in Fig. \ref{fig:ENF090}a, and in the corresponding damage profile in Fig. \ref{fig:ENF090}c with  standard FEM, whereas no traction oscillations are seen with the stabilized FEM. 

\subsection{Fixed Ratio mixed mode test}\label{FRMM delamination test}

\begin{figure}[ht]
\centering
\includegraphics[width=1\textwidth]{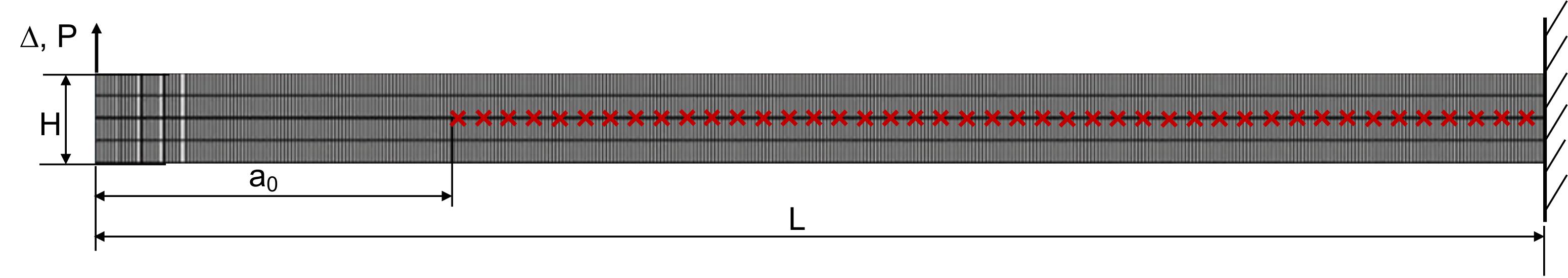}
\caption{Geometry and boundary conditions for the fixed ratio mixed mode test. The dimensions are: L = 50 mm, H = 3.1 mm and a$_{0}$ = 35 mm.}
\label{fig:FRMMTestSetup}
\end{figure}

\begin{figure}[ht]
 \centering
 \begin{subfigure}[b]{0.4\linewidth}
   \centering\includegraphics[width=\textwidth]{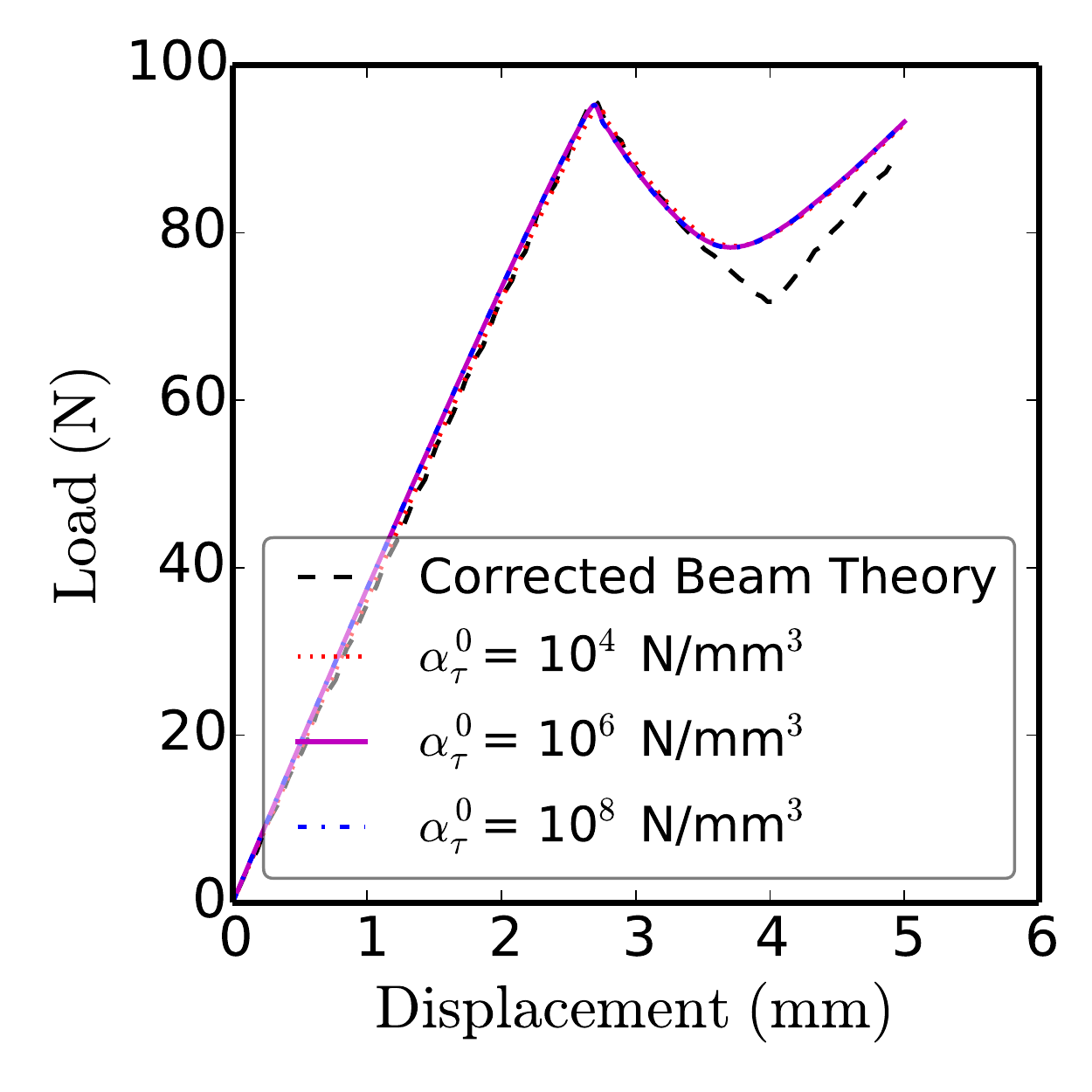}
   \caption{}
       \label{fig:standardFRMM00loaddisp}
 \end{subfigure}%
 ~
 \begin{subfigure}[b]{0.4\linewidth}
   \centering\includegraphics[width=\textwidth]{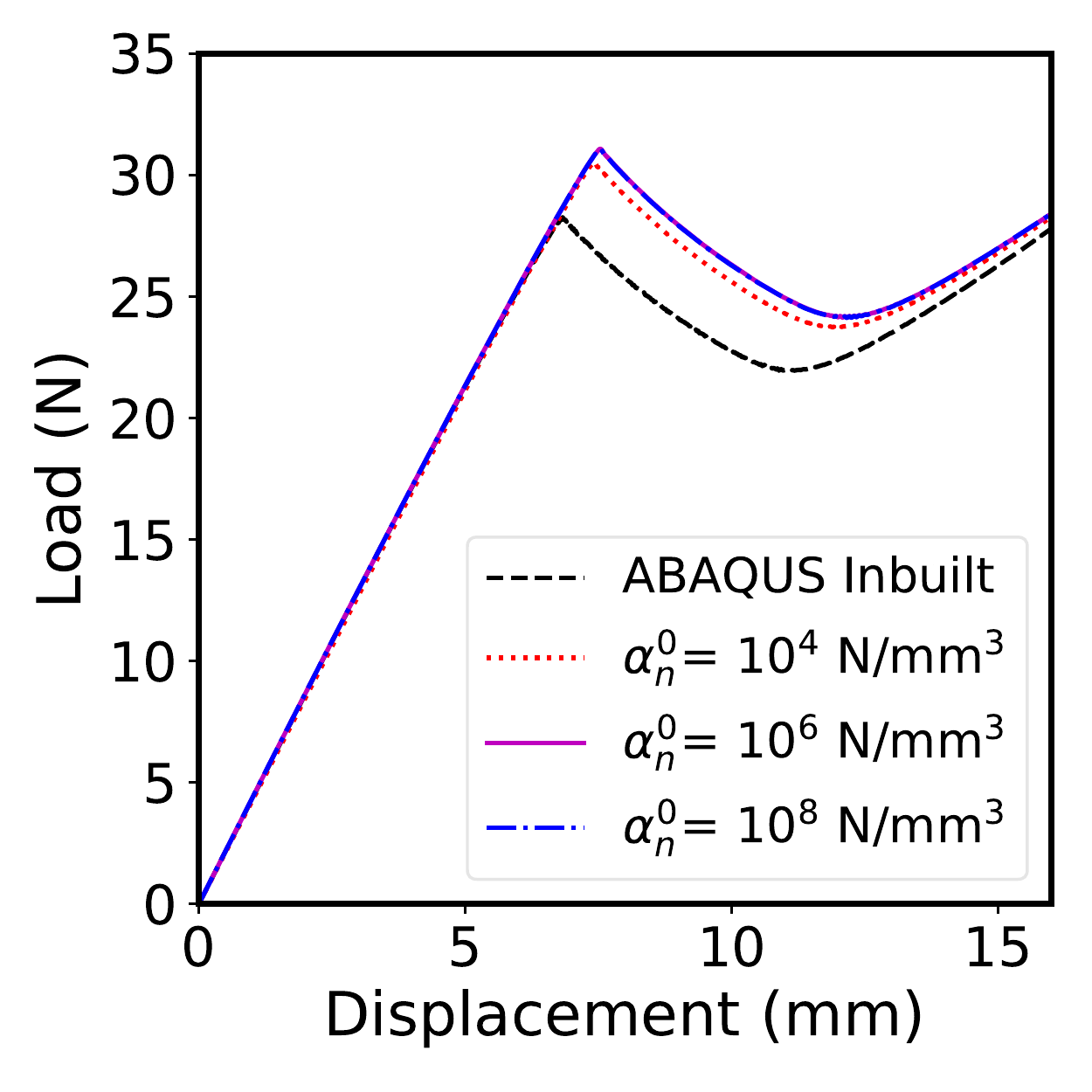}
   \caption{}
       \label{fig:stabilizedFRMM00loaddisp}
 \end{subfigure}
 \caption{Load versus displacement curves for the fixed ratio mixed mode test obtained from the stabilized FEM: (a) [0/0] orientation and (b) [0/90] orientation.} \label{fig:FRMMstblloaddisp}
\end{figure}

We also investigated the ability of the stabilized FEM in recovering normal and tangential traction fields during mixed-mode delamination crack growth using the fixed ratio mixed mode (FRMM) test. The specimen geometry and test set-up along with the finite element mesh are shown in Fig. \ref{fig:FRMMTestSetup}. To initiate the delamination process, a pre-crack is placed at the left end and vertical displacement is applied only at the upper corner node. The fixed boundary condition is applied at the right end of the beam. 
In Fig. \ref{fig:FRMMstblloaddisp}a, we compare the load--displacement curves from the stabilized FEM with the corrected beam theory solution from \citep{harper2008cohesive} for [0/0] laminate, due to the lack of experimental data. In Fig. \ref{fig:FRMMstblloaddisp}b, we compare the load--displacement curves from the stabilized FEM with the ABAQUS inbuilt cohesive elements for [0/90] laminate, due to the lack of experimental data or analytical solutions. Because the ABAQUS inbuilt cohesive elements use a different mixed-mode criteria proposed in \citep{camanho2003numerical}, the predicted peaks loads and softening portions of the load--displacements are different in Fig. \ref{fig:FRMMstblloaddisp}b. However, by choosing different cohesive strengths with different mixed-mode criteria it is possible to match the load--displacement curves obtained from stabilized FEM and inbuilt ABAQUS elements. 

\subsubsection{Traction at [0/0] laminate interface} 

\begin{figure}[ht]
\centering
\includegraphics[width=1\textwidth]{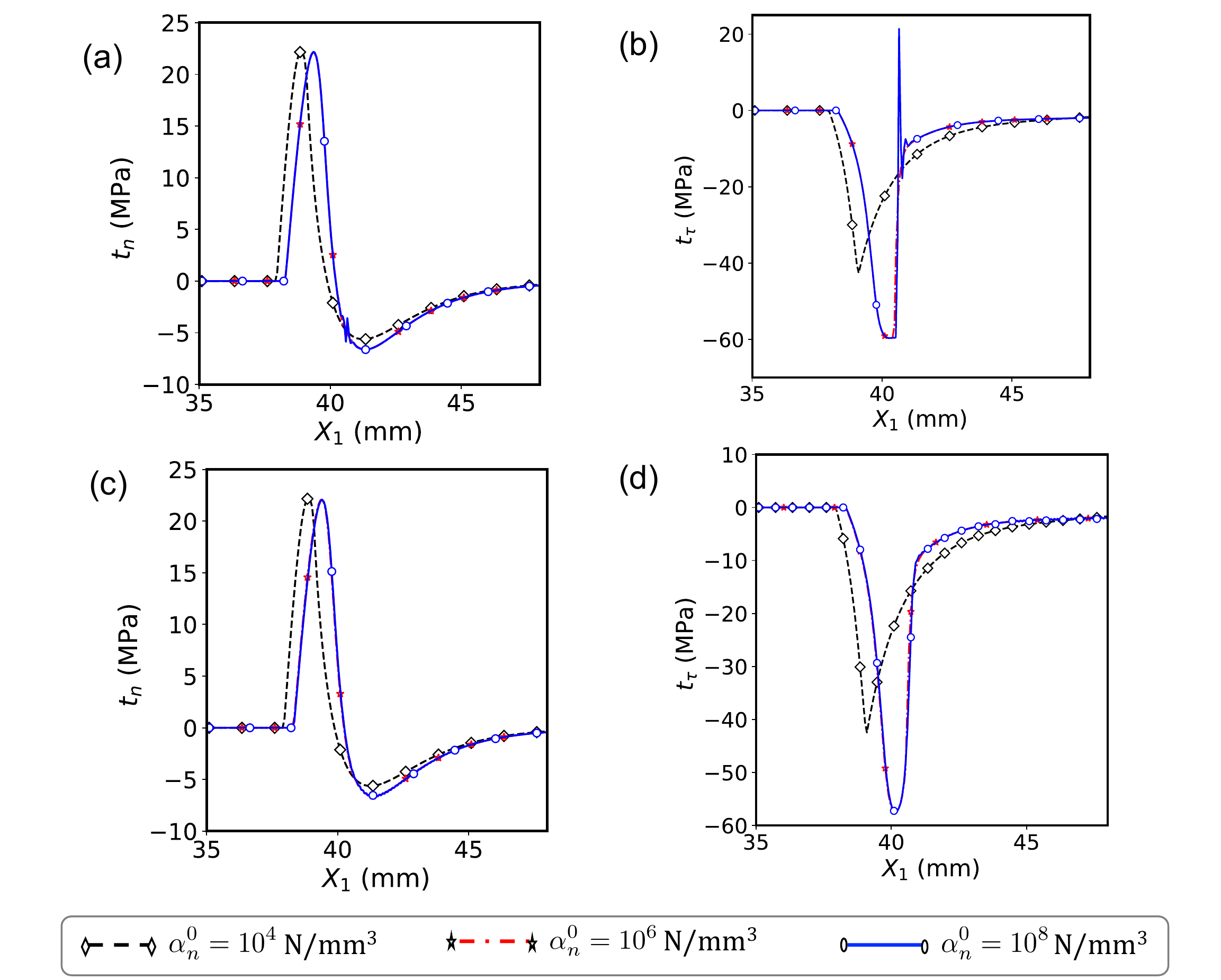}
\caption{(a) Normal and tangential traction fields versus interface length from the fixed ratio mixed mode test ([0/0] ply orientation) for different cohesive stiffness values: (a), (b) standard FEM and (c), (d) stabilized FEM.} \label{fig:FRMM00}
\end{figure}

In Figs. \ref{fig:FRMM00}a and \ref{fig:FRMM00}b, we show the normal and tangential traction profiles for [0/0] laminate from the standard FEM. We find that for the large cohesive stiffness $\alpha_n^0 = \alpha_{\tau}^0= 10^{8}$ N/mm$^{3}$, large amplitude oscillations appear in the tangential traction, but a few small amplitude oscillation appears in the normal traction profile. The stabilized FEM alleviates oscillations in both normal and tangential traction fields, as evident from Figs. \ref{fig:FRMM00}c and \ref{fig:FRMM00}d. Notably, for the smaller stiffness value of $10^{4}$ N/mm$^{3}$, the peak tangential traction value is considerably smaller than the other two cohesive stiffness cases (i.e., $10^{6}$ N/mm$^{3}$ and $10^{8}$ N/mm$^{3}$) from both methods, which also happens with ABAQUS in-built cohesive elements. This suggests that it is important to take the cohesive stiffness large enough to accurately recover the interface  traction for mixed mode loading. 

\subsubsection{Traction at [0/90] laminate interface}

\begin{figure}[ht]
\centering
\includegraphics[width=1\textwidth]{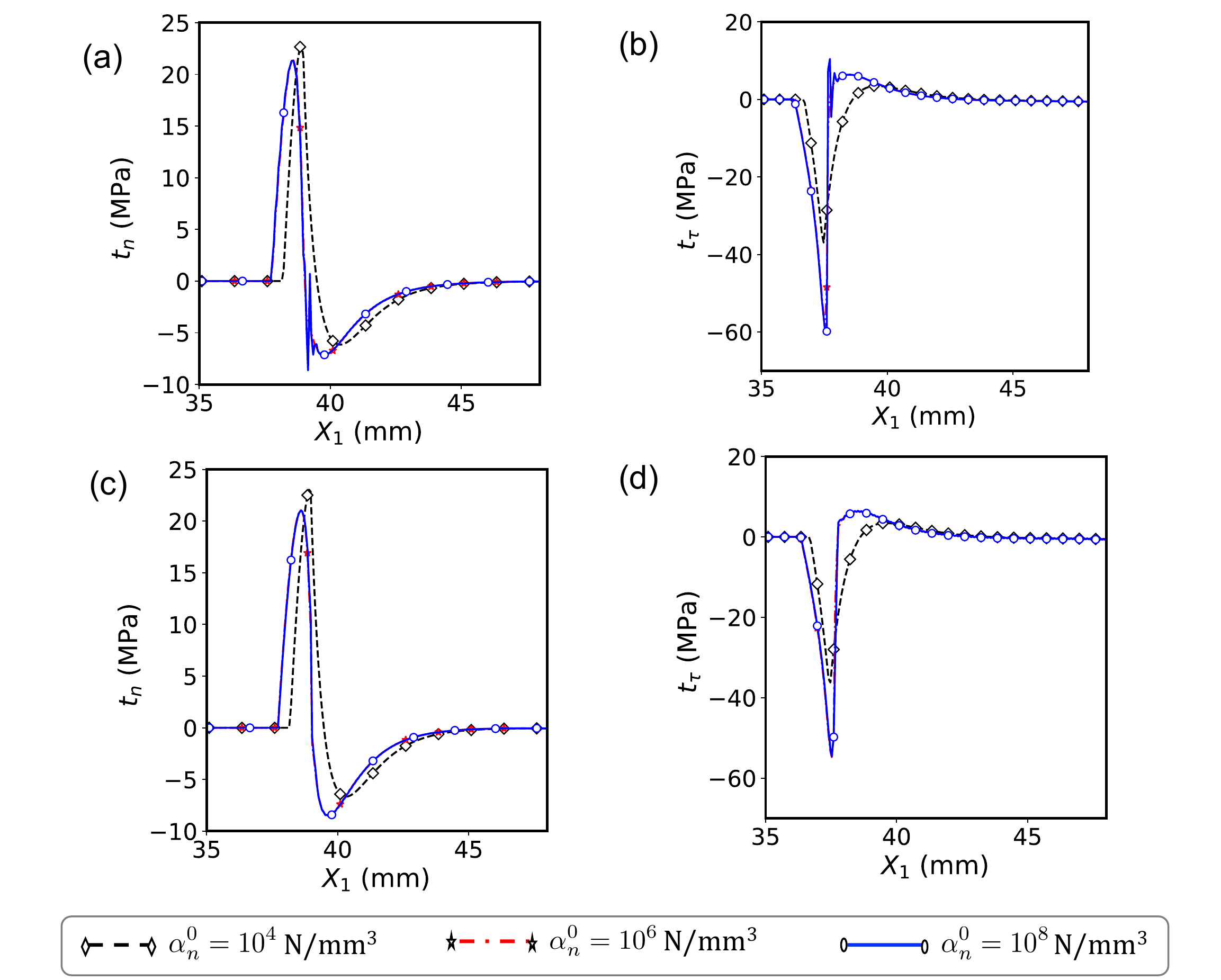}
\caption{Normal and tangential traction fields versus interface length from the fixed ratio mixed mode test ([0/90] ply orientation) for different cohesive stiffness values: (a), (b) standard FEM and (c), (d) stabilized FEM.} \label{fig:FRMM090}
\end{figure}

In Figs. \ref{fig:FRMM090}a and \ref{fig:FRMM090}b, we show the normal and tangential traction profiles for [0/90] laminate from the standard FEM. We find that for $\alpha_n^0 = \alpha_{\tau}^0= 10^{8}$ N/mm$^{3}$, large amplitude oscillations appear in both normal and tangential traction profiles. The stabilized FEM alleviates these oscillations, as evident from Figs. \ref{fig:FRMM090}c and \ref{fig:FRMM090}d. We also notice for the smaller stiffness value of $10^{4}$ N/mm$^{3}$, the peak tangential traction value is considerably smaller than the other two cohesive stiffness cases (i.e., $10^{6}$ N/mm$^{3}$ and $10^{8}$ N/mm$^{3}$). Once again, this indicates the importance of taking a larger value for cohesive stiffness to accurately recover the interface traction. 

\section{Conclusion}\label{sec:Conc}
In this paper, we compared and contrasted the performance of standard and stabilized finite element methods for composite delamination analysis using cohesive elements. First, we illustrated through constant strain patch tests that numerical instability with standard FEM causes spurious oscillations in the interface traction fields at both similar and dissimilar composite material interfaces. We demonstrated that perturbed, semi-structured and potentially unstructured meshes can aggravate the numerical instability with standard FEM, especially in anisotropic CZMs where the contrast between normal and tangential stiffness is high. For constant strain patch tests, we then illustrated the ability of a stabilized method inspired by the weighted Nitsche approach in alleviating spurious traction oscillations at the delamination interface.

We next examined the traction and damage profiles recovered from the standard and stabilized methods using delamination tests (\emph{i.e.,} DCB, ENF and FRMM tests). Once again, spurious oscillations were observed in normal and tangential traction fields for a moderately large choice of the cohesive stiffness parameter (\textit{i. e.,} three orders of magnitude larger than the bulk stiffness). However, we found that the load--displacement curves obtained from both standard and stabilized FEM match well with theory and experiments, implying that the numerical instability observed in the standard method does not corrupt it's load predictions. The key point is that the numerical instability seems to only corrupt the traction fields recovered from the standard FEM when the cohesive stiffness parameter is taken to be moderately large. However, simply taking a small cohesive stiffness parameter is not appropriate because it diminishes the accuracy of the traction field, especially under mixed-mode loading, where the interface damage occurs before the maximum cohesive strength is reached. In contrast, our stabilized FEM ensures stability even with large values of cohesive stiffness, and enables accurate recovery of the interface traction field. 
We found that standard FEM with structured uniform meshes is less prone to instability in the three delamination tests. Typically, we only find isolated oscillations in normal and tangential traction profiles near the crack tip region, where the traction and damage fields varies sharply. However, the standard FEM with perturbed meshes exhibits spurious oscillations in compressive/contact regions of the delamination interface, although they are smaller in amplitude. 

In conclusion, with the standard FEM the choice of cohesive stiffness can be non-trivial and involves a precarious trade-off -- prescribing a value that is larger than necessary results in numerical instabilities, whereas prescribing a value that is smaller than necessary results in inaccurate traction recovery. A ``correct" choice of cohesive stiffness is not always apparent and can only be found by conducting sensitivity studies based on 1D model estimates. In contrast, the proposed stabilized FEM is robust and stable for a wide range of cohesive stiffness values. Furthermore, as discussed in Ghosh \textit{et al.} \citep{ghosh2019stabilized} the stabilized FEM does not increase the computational cost compared to the standard FEM, so it is advantageous for fracture and composite delamination analysis using cohesive/interface elements.

\section*{Acknowledgements}
\thispagestyle{empty}
Gourab Ghosh and Ravindra Duddu gratefully acknowledge the financial support of the Office of Naval Research -- award
\#N0014-17-12040 (Program Officer: Mr. William Nickerson). Chandrasekhar Annavarapu would like to gratefully acknowledge the funding received through the New Faculty Initiation Grant (NFIG) at Indian Institute of Technology Madras. We also thank Prof. Caglar Oskay at Vanderbilt University for his helpful comments on composite delamination analysis.

\bibliographystyle{unsrt}
\bibliography{bibCZM}
\biboptions{sort&compress}

\end{document}